\documentclass[aps,superscriptaddress]{revtex4}  
\usepackage[switch]{lineno}
\usepackage{amsmath}
\usepackage{float}
\usepackage{amssymb}
\usepackage{empheq}
\usepackage[parfill]{parskip}
\usepackage[active]{srcltx}
\usepackage{color}
\usepackage{array}
\usepackage{booktabs}
\usepackage{amsfonts}
\usepackage{dsfont}
\usepackage{graphicx}

\begin{document}

\setlength{\parindent}{0.5cm}

\title{Swarmalators with delayed interactions}

\author{Nicholas Blum}
\affiliation{California Polytechnic State University, San Luis Obispo, CA 93407}
\email{kevin.p.okeeffe@gmail.com, okogan@calpoly.edu}

\author{Andre Li}
\affiliation{California Polytechnic State University, East Bay, CA 94542}

\author{Kevin O'Keeffe}
\affiliation{Senseable City Lab, Massachusetts Institute of Technology, Cambridge, MA 02139} 

\author{Oleg Kogan}
\affiliation{California Polytechnic State University, San Luis Obispo, CA 93407}

\begin{abstract}

We investigate the effects of delayed interactions in a population of ``swarmalators", generalizations of phase oscillators that both synchronize in time and swarm through space. We discover two steady collective states: a state in which swarmalators are essentially motionless in a disk arranged in a pseudo-crystalline order, and a boiling state in which the swarmalators again form a disk, but now the swarmalators near the boundary perform boiling-like convective motions. These states are reminiscent of the beating clusters seen in photoactivated colloids and the living crystals of starfish embryos.\\

\end{abstract}

\maketitle

\section{Introduction}
Swarmalators are generalizations of phase oscillators that swarm around in space as they synchronize in time \cite{o2017oscillators}. They are intended as prototypes for the many systems in which sync and swarming co-occur and interact, such as biological microswimmers \cite{yang2008cooperation,riedel2005self,quillen2021metachronal,peshkov2021wiggling,peshkov2022synchronized,tamm1975role,verberck2022wavy,belovs2017synchronized}, forced colloids \cite{yan2012linking,hwang2020cooperative,zhang2020reconfigurable,bricard2015emergent,zhang2021persistence,manna2021chemical,li2018spatiotemporal,chaudhary2014reconfigurable}, magnetic domain walls \cite{hrabec2018velocity,haltz2021domain}, robotic swarms \cite{barcis2019robots,barcis2020sandsbots,schilcher2021swarmalators,monaco2020cognitive,hadzic2022bayesian,rinner2021multidrone,gardi2022microrobot,origane2022wave} and embryonic cells of starfish \cite{tan2021development} and zebrafish \cite{petrungaro2019information}.

Research on swarmalators is rising. Tanaka et. al.~began the endeavour by introducing a universal model of chemotactic oscillators with rich dynamics \cite{tanaka2007general,iwasa2010hierarchical,iwasa2012various,iwasa2017mechanism}. Later O’Keeffe, Hong, and Strogatz studied a mobile generalization of the Kuramoto model \cite{o2017oscillators}. This swarmalator model is currently being further studied. The effects of phase noise \cite{hong2018active}, local coupling \cite{lee2021collective,jimenez2020oscillatory,schilcher2021swarmalators,japon2022intercellular}, external forcing \cite{lizarraga2020synchronization}, geometric confinement \cite{o2022collective,yoon2022sync,o2022swarmalators}, mixed sign interactions \cite{mclennan2020emergent,hong2021coupling,sar2022swarmalators,sar2022swarmalators}, and finite population sizes  \cite{o2018ring} have been studied. The well posedness of weak and strong solutions to swarmalator models have also been addressed \cite{ha2021mean,ha2019emergent,degond2022topological}. Reviews and potential application of swarmalators are provided here \cite{sar2022dynamics,o2019review}. Mobile oscillators, where oscillators' movements affect their phases but not conversely, have also been studied \cite{fujiwara2011synchronization,uriu2013dynamics,levis2017synchronization,majhi2019emergence}.

This paper is about swarmalators with delayed interactions. Delays are, in this context, largely unstudied, although they occur commonly in Nature and technology. In the case of microswimmers, the inter-swimmer coupling is mediated by the surrounding fluid and is therefore non-instantaneous.  Delays are also an important factor to consider in embryonic development.  Here, they are a well established feature of gene expression and are believed to play a key role in how cells, organs, and other agglomerations attain their shapes \cite{petrungaro2019information}.  The authors of \cite{petrungaro2019information} write: `` Even though cell coupling is local, involving cells which are in direct contact, cells require some time to synthesize and transport the membrane ligands and receptors to their surface. Also, cells need time to integrate received information to its internal gene expression dynamics, for example, by producing transcription factors. Each of these reaction processes takes a different time to be completed, and these times depend on cell type and cell state.''  The continue: ``This time delay might be relevant for cell coupling because what cells acquire at the present time is the information of surrounding cells some time ago. Thus, inherent delays in cell coupling are key to understanding information flow in biological tissues.''  Time delay is also relevant to robotic swarms where digital communication comes with unavoidable lags and may affect both communication of the spatial or internal state of robots.  

In short, delays are important for a broad class of swarmalators;  in some cases delay affects the communication of internal state of particles, and in others it affects the communication of both the internal and spatial state.  He we aim to take a first step in understanding delay-coupled swarmalators theoretically, so we will focus on delays in just the internal state of the original swarmalator model \cite{o2017oscillators}. This model is a natural first case-study because it captures the behaviors of many natural swarmalators \cite{quillen2021metachronal,peshkov2022synchronized,zhang2020reconfigurable} yet is simple enough to analyze.

\section{Swarmalators with delay}
We will introduce time delay into the swarmalator model proposed by O’Keeffe, Hong, and Strogatz (OHS) \cite{o2017oscillators}.  The equations describing the dynamics of such delayed swarmalators read
\begin{align}
& \dot{\mathbf{x}}_i = \mathbf{v}_i +   \frac{1}{N} \sum_{ j \neq i}^N \Bigg[ \frac{\mathbf{x}_j - \mathbf{x}_i}{|\mathbf{x}_j - \mathbf{x}_i|} \Big( 1 + J \cos(\theta_j(t-\tau) - \theta_i(t))  \Big)  -   \frac{\mathbf{x}_j - \mathbf{x}_i}{ | \mathbf{x}_j - \mathbf{x}_i|^2}\Bigg]   \label{eq:position_eq} \\
& \dot{\theta_i} = \omega_i +  \frac{K}{N} \sum_{j \neq i}^N \frac{ \sin(\theta_j(t-\tau) - \theta_i(t))}{ |\mathbf{x}_j - \mathbf{x}_i| }. \label{eq:theta_eq}
\end{align}
Here $N$ is the number of particles.  All the spatial coordinates are evaluated without delay, at time $t$.  The first term in Eq.~(\ref{eq:position_eq}) represents attraction - it causes the velocity of particle $j$ to be directed towards particle $i$ and vice versa. The parameter $J$ controls the tendency of this attractive term to depend on internal phases; when $J=0$ the attraction is independent of internal phases.  In order for the first term to be attractive, $|J|$ must be less than $1$.  The attractive term has a magnitude that's independent of the particle separation, i.e. it represents an all-to-all attraction (which could also be called mean field attraction) that is commonly used.  For example, the same is done with phases in the Kuramoto model.

The second term in Eq.~(\ref{eq:position_eq}) is a short-range repulsion - it causes the velocity of particle $i$ to be directed away from particle $j$, but this term decays away with distance.  It is intended to prevent clumping of all particles at one point.  The form of the model is motivated in \cite{o2017oscillators}.  

Eq.~(\ref{eq:theta_eq}) describes dynamics of internal phases.  If $\theta_j$ is lagging behind $\theta_i$, the $i-j$ term in the sum contributes to the velocity of $\theta_i$ that tends to bring $\theta_i$ closer to $\theta_j$.  In other words, oscillator $j$ ``pulls'' the phase of oscillator $i$ closer to it.  This is the usual Kuramoto interaction.  The parameter $K$ is an overall scaling factor for the strength of phase attraction (positive $K$) or repulsion (negative $K$).  Here the strength of the interaction depends on the distance - oscillators that are closer in physical space will experience stronger tendency to align or counter-align their phases with their close neighbors.  

Thus, the picture is this: the phase dynamics affects the strength of spatial attraction, while the spatial position of particles affects the strength of phase interaction.  

The new addition in this work is the time delay in phase dependence.  The particle $i$ at time $t$ responds to the phase of the particle $j$ as it was a time $\tau$ ago - at time $t-\tau$.  In this work, we only add this effect to the phase dynamics.  Physically, the phase represents the internal state - for example, the phase of a gene expression cycle.  Communication of such internal variable often takes place via chemical signals, which is a type of interaction that is much slower than the interaction that communicates positions of objects in physical space \cite{petrungaro2019information}. 

We investigate the role of delay in this model.  OHS discovered that the system can be found in one of the five collective states in the absence of a delay.  In the present paper we work mostly in the region of $(J,K)$ space that in this delay-free model corresponds to what OHS called the ``active phase wave''.  The swarmalators in this state move in circles around the center of the annulus-shaped cluster - some move clockwise, and some counterclockwise, while the internal phases change as they move around the center of the annulus.  It is in this region of the $(J,K)$ space is where we found the interesting collective behavior induced by a delay.  It is possible that other new behaviors take place in other regions of $(J,K)$ space, but this would be a subject for a future work.  

This plan of this paper is as follows.  In Section \ref{sec:num_results} numerical results are presented.  Two new collective states are presented in Subsections \ref{sec:pseudo-static_num} and \ref{sec:boiling_num} respectively.  Subsection \ref{sec:transitions} describes the dynamical phase transition between these states, as well as properties of long-time behavior of transient oscillations.  Theoretical treatment is presented Section \ref{sec:Theory}.  We compare theoretical predictions with numerical results there.  We summarize in Section \ref{sec:discission}.

\section{Two new types of collective behavior - numerical results}
\label{sec:num_results}
We begin by presenting a phenomenon that occurs at sufficiently large $\tau$.  The meaning of ``sufficiently large'' will be made precise in Section \ref{sec:transitions}.  
\subsection{Pseudo-static quasi-crystal}
\label{sec:pseudo-static_num}
\begin{figure}[H]
\center \includegraphics[width=6in]{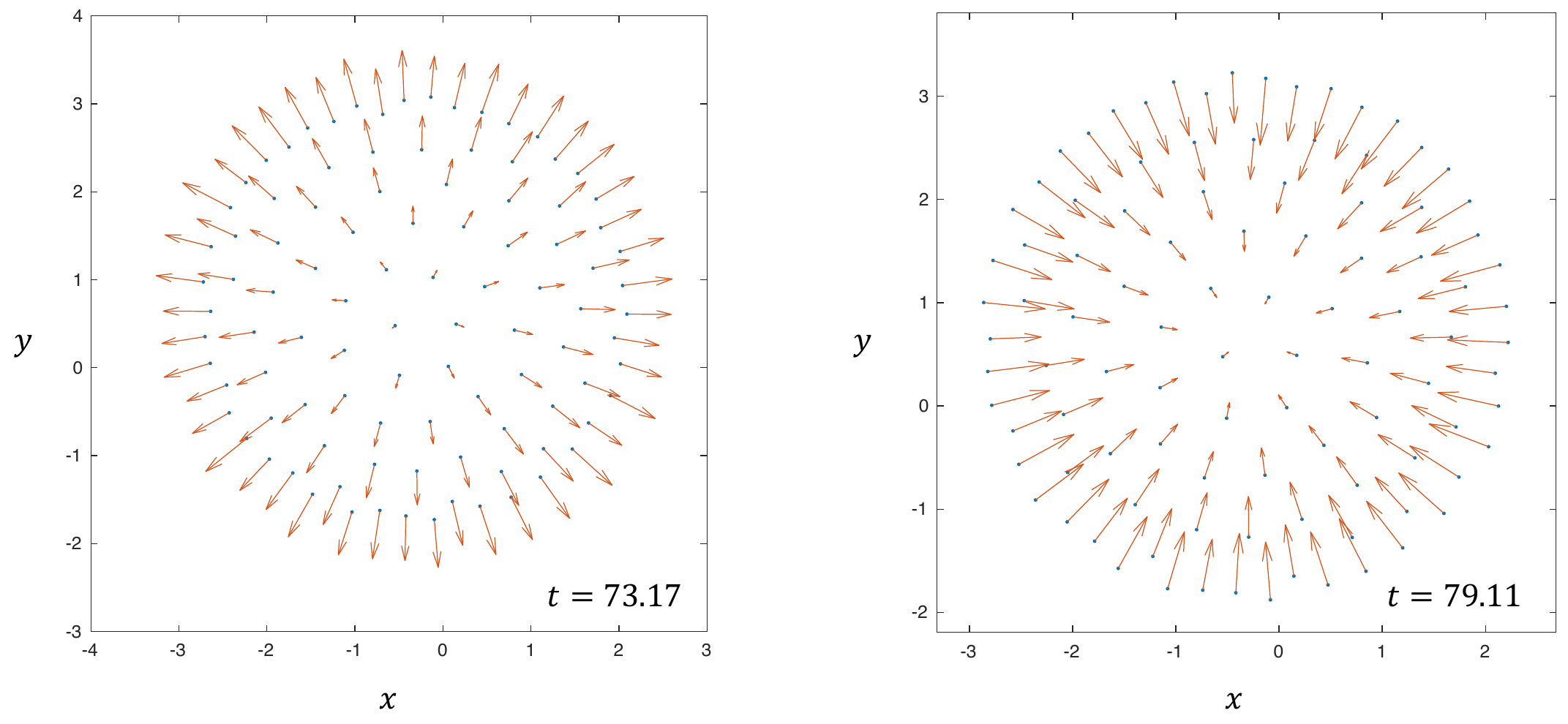}
\caption{Velocity vectors and particle positions at two instants of time.  These times are at latter stages of oscillations (see Fig.~\ref{fig:avg_R_and_avg_v_breathing} for corresponding speeds and the average radius (not maximum radius) at these times). $(N,J,K,\tau)=(100,1,-0.75,8)$. }
\label{fig:vectors_breathing}
\end{figure}

\begin{figure}[H]
\center \includegraphics[width=7in]{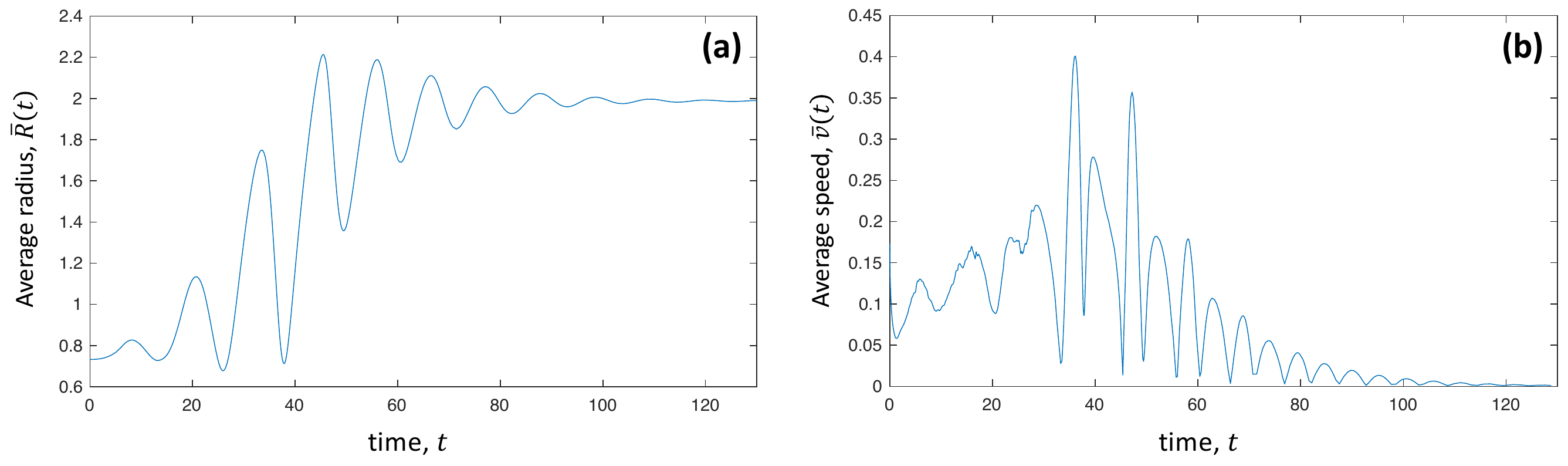}
\caption{(a) $\overline{R}(t)$.  (b) $\overline{v}(t)$.  The parameters are $(N,J,K,\tau )=(100,1,-0.75,8)$}
\label{fig:avg_R_and_avg_v_breathing}
\end{figure}

We placed particles at random positions \textbf{x} within a square with corners at $(\pm 1, \pm 1)$, and assigned initial phases $\theta$ uniformly at random from $[0,2 \pi]$.  Following an initial complicated transient, when particles quickly organize into a nearly-circular disk, it enters into a coherent, synchronized collective motion characterized by decaying oscillations of the radius.  Fig.~\ref{fig:vectors_breathing} demonstrates velocity vectors of particles at two snapshots in time - one during expansion, and another during contraction - while Fig.~\ref{fig:avg_R_and_avg_v_breathing}(a) demonstrates oscillations of the average radius of the cluster, $\overline{R}(t)=\frac{1}{N}\sum_{n=1}^N R_n(t)$.   Naturally, the average velocity and the average speed of particles also exhibit oscillations.  Fig.~\ref{fig:avg_R_and_avg_v_breathing}(b) depicts the average speed $\overline{|v|}(t) = \frac{1}{N} \sum_{n=1}^N |v|_n(t)$.
%
%
We will refer to this collective behavior as ``breathing'' of the cluster.  Because the oscillations during the  breathing decay, this stage of the system dynamics can be thought of as the longer portion of the transient.  At earlier times, the transient is more complicated, and does not result in breathing motion.  
The dynamics at very earlier times is complex.  The first few breaths are also complicated - they are not purely radial, and can be accompanied by other types of dynamics - including particle rearrangements and time-averaged expansion of the cluster (this is why the speed doesn't go to zero at maximum and minimum radius).  Eventually, breathing motion becomes simpler - it consists of only radial oscillations around the infinite-time equilibrium radius value, and there are no particle rearrangements in this latter stage.  In Fig.~\ref{fig:avg_R_and_avg_v_breathing} this happens around $t=80$.

\begin{figure}[H]
\center \includegraphics[width=6.9in]{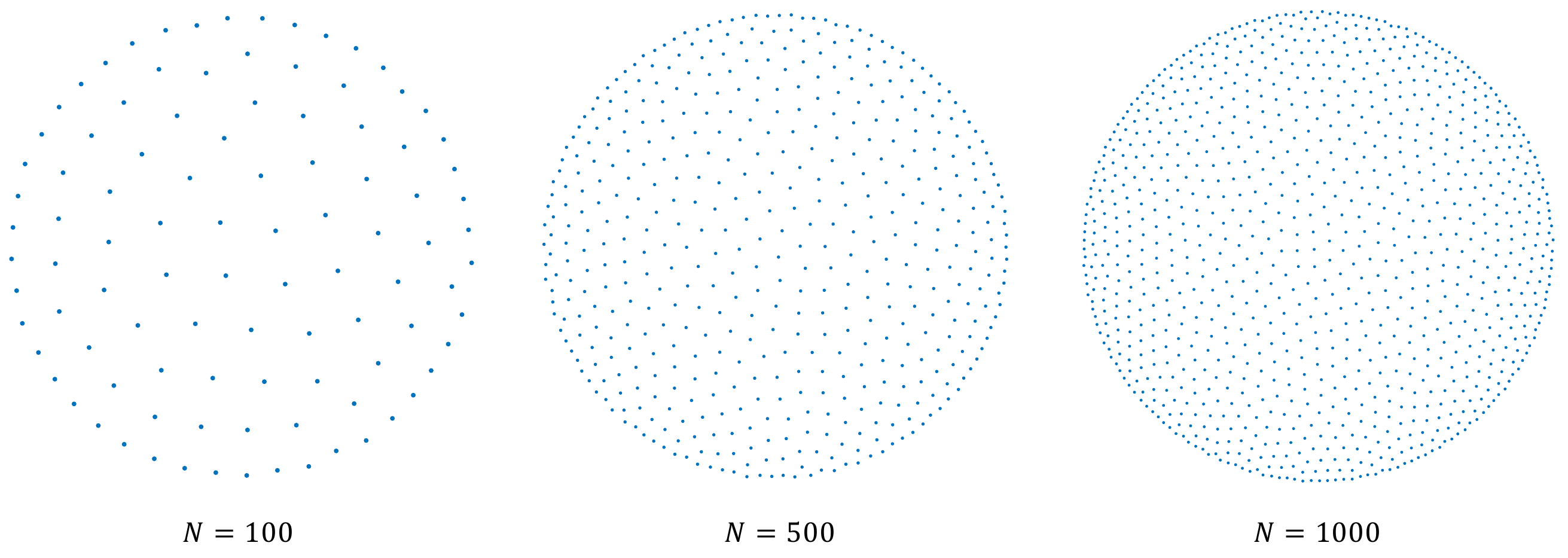}
\caption{$N=100$ (left), $N=500$ (center) and $N=1000$ (right) particle systems after the breathing transient.  The values of $\tau$ were $1.5 \tau_c$, where $\tau_c$ is a critical delay at which the long-time behavior becomes boiling motion, as described in the next Subsection (see Fig.~\ref{fig:tauc_vs_N} for values of $\tau_c$ at different $N$). Here $J=1$, $K=-0.7$.}
\label{fig:static}
\end{figure}

\begin{figure}[H]
\center \includegraphics[width=3in]{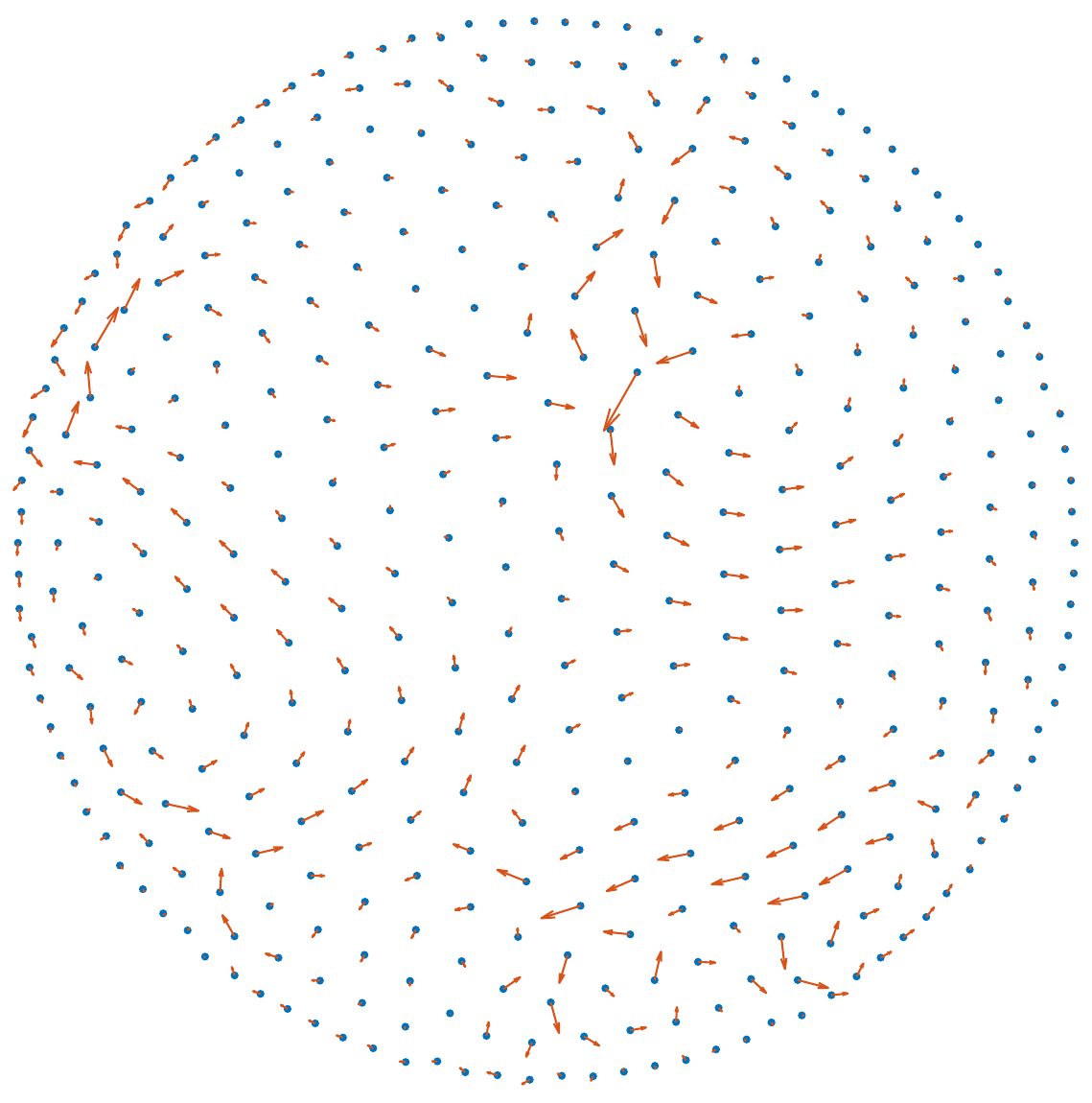}
\caption{Creeping particle motion for $t\approx 5000$, which is approximately $10$ times the time at which the breathing ceased, as defined above.  Here $N=400$, and the values of $\tau$ was $1.5 \tau_c$ (see Fig.~\ref{fig:tauc_vs_N} for values of $\tau_c$), $J=1$, and $K=-0.7$. The vectors have been automatically re-scaled to be visible.  Thus, while the arrows appear to have the length comparable to those in Fig.~\ref{fig:vectors_breathing}, this is because of the up-scaling.}
\label{fig:latter_vectors}
\end{figure}

\begin{figure}[H]
\center \includegraphics[width=4in]{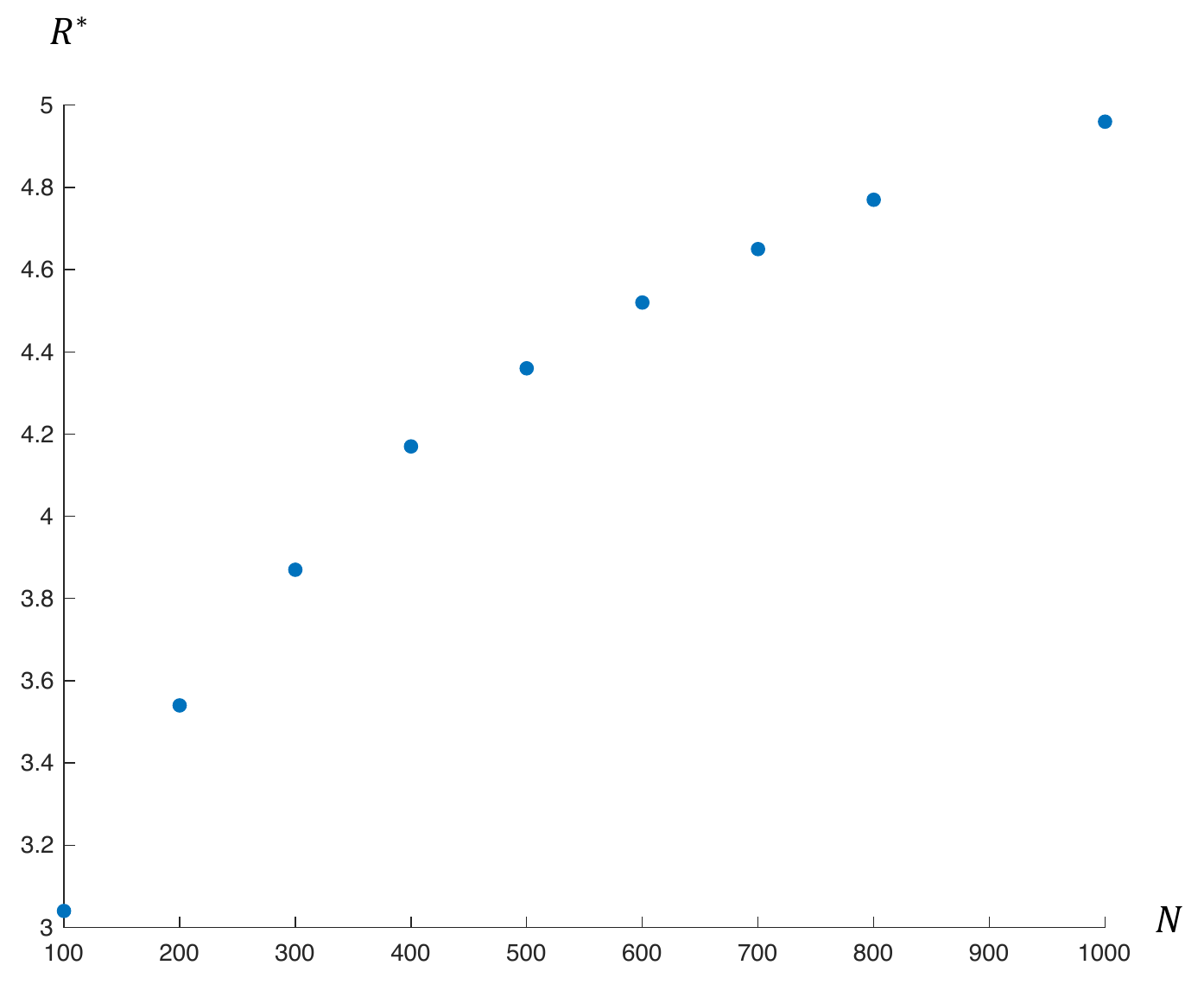}
\caption{$(J,K)=(1,-0.7)$.  The $\tau$ value was not fixed;  rather the fixed quantity was the fraction that $\tau$ represented above the critical $\tau_c$ (see Subsection \ref{sec:transitions}), which itself is a function of $N$.}
\label{fig:radius_N_dependence}
\end{figure}

After the breathing transient dies down, it appears that a static pseudo-crystal is formed.  Examples of these pseudo-crystals are shown in Fig.~\ref{fig:static} for three system sizes.  Note that the radii of these  pseudo-crystals depend on $N$, i.e. the radii of the three clusters in Fig.~\ref{fig:static} are not equal (see Fig. \ref{fig:radius_N_dependence}) - they have been scaled in Fig.~\ref{fig:static}.  But the inter-particle spacing relative to the cluster radius clearly decreases with larger $N$. 
However, a careful examination of the tail of $\overline{R}(t)$ plot demonstrates that there is some residual motion left.  This is seen on the logarithmic plot below.
\begin{figure}[H]
\center \includegraphics[width=5.8in]{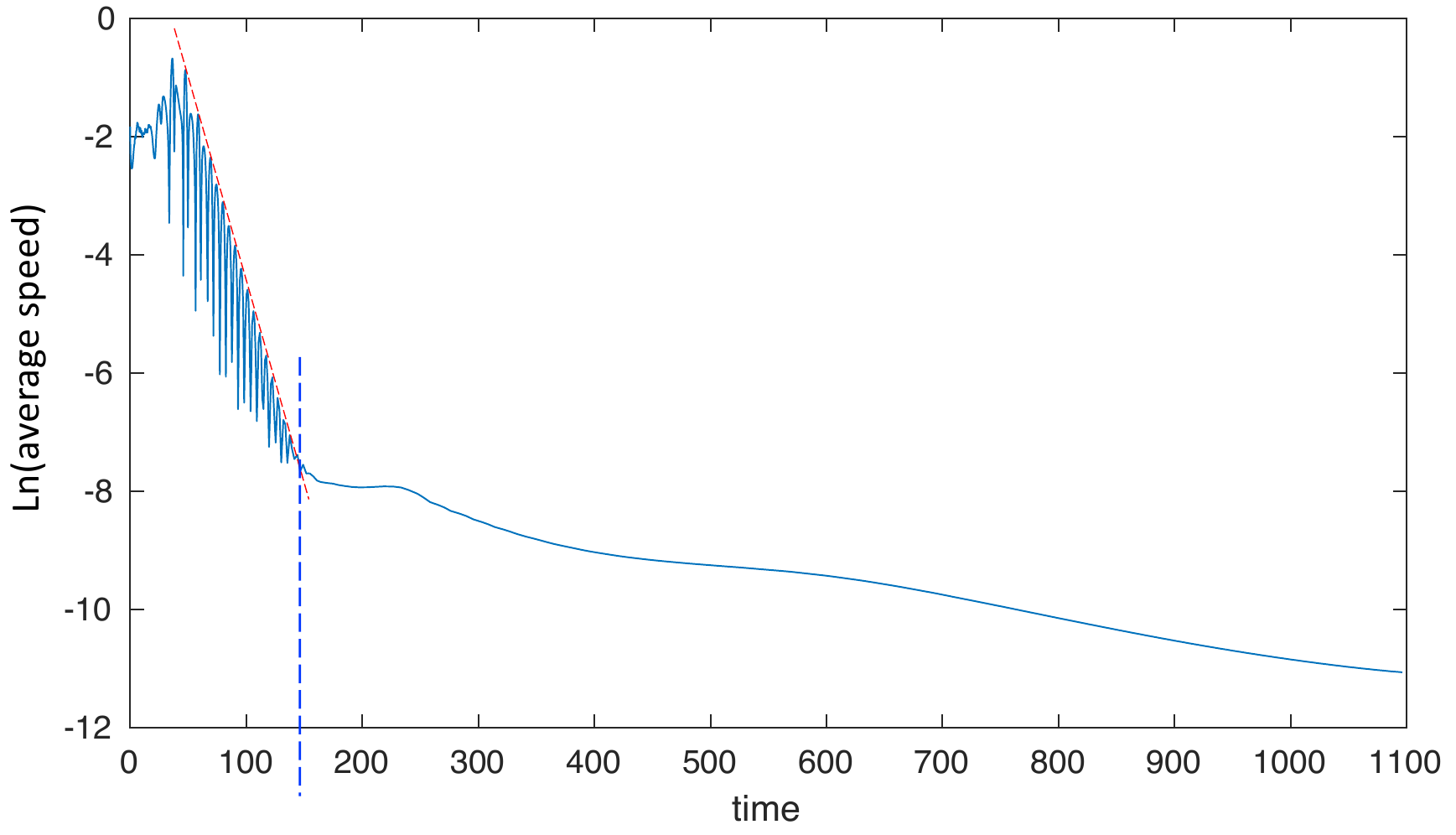}
\caption{$(N,J,K,\tau)=(100,1,-0.75,8)$. Natural logarithm of the average speed versus time.  A transition from decaying oscillatory motion to creeping motion is very clear.}
\label{fig:residual}
\end{figure}
Around $t=150$ we clearly see that breathing motion gives way to a different type of motion with very small velocities (we can call it creeping motion).  The magnitude of these velocities continues to decay with time, but much slower than during the breathing stage.  We can define the transition to this creeping motion as an intersection of the straight-line fit on the logarithmic plot to the envelope of $\overline{|v|}(t)$ during breathing (red dashed line in Fig.~\ref{fig:residual}) with the function itself.  There is no single defining feature of this post-breathing velocity pattern - its character changes with time and with respect to parameters.  The only definite feature of these post-breathing creeping dynamics is that it is rather disordered.  We give an example of such a pattern in Fig.~\ref{fig:latter_vectors}.  Additional examples can be found in Figs.~\ref{fig:N100_mousetail} and \ref{fig:N400_mousetail} in Appendix \ref{sec:Details}.

\begin{figure}[H]
\center \includegraphics[width=4.8in]{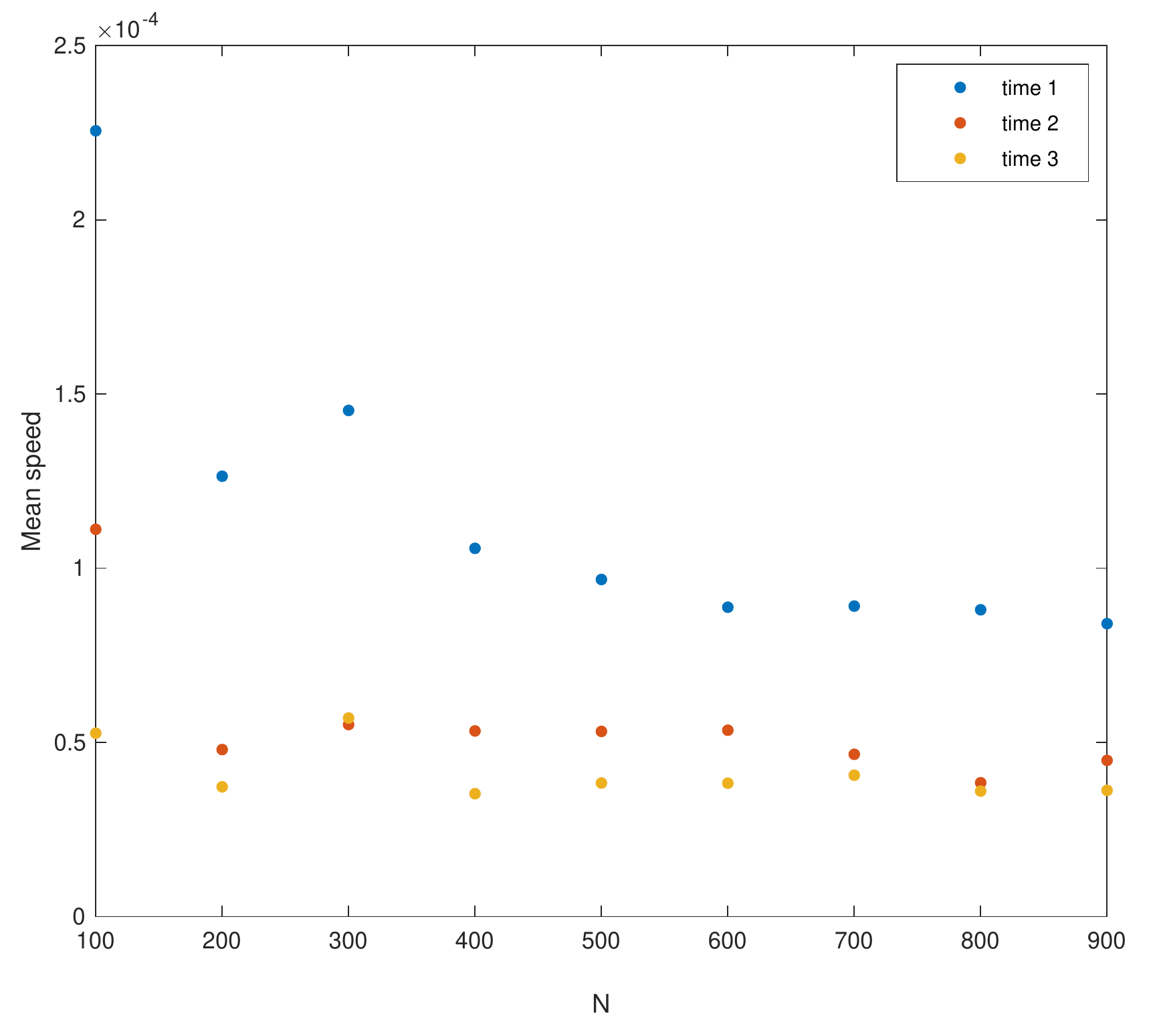}
\caption{$(J,K)=(1,-0.7\color{black})$. Time 1 is immediately after the end of the breathing motion, time 2 around $500$ time units after the end of the breathing motion, and time 3 around $1000$ time units after the end of the breathing motion.}
\label{fig:no_N_dependence}
\end{figure}

There is no indication, given the range of our computational capabilities, that the post-breathing creeping motion is a finite size effect.  We come to this conclusion by measuring the dependence of the average speed on $N$ at three instants of time that follow the breathing.  
In Fig.~\ref{fig:no_N_dependence} we plot the average speed versus $N$ measured at three instants in time: the first, labeled ``time 1'' is immediately after the end of the breathing motion as just defined.   The second, or ``time 2'' is around $500$ time units after end of the breathing motion, and the third, or ``time 3'' is $1000$ time units after the end of the breathing motion.  The data in Fig.~\ref{fig:no_N_dependence} suggests that there is no indication (at least in the range of $N$s that were studied) that the long-time average speed decreases with increasing system size.  

Other properties do exhibit $N$-dependence.  For example, Fig.~\ref{fig:radius_N_dependence} clearly demonstrates that there is $N$ dependence of the radius of the cluster $R^*$ after breathing. 

So far we discussed the spatial aspect of the dynamics.  The following plot depicts the dynamics of internal phases.  We see again the early transient, the longer portion of the transient coincident with the breathing stage, followed by the long-time behavior of uniform growth of internal phases all at the same rate.  During the longer stage of the transient the phases undergo a series of plateaus followed by short and rapid collective phase slips.  Note that at this time all the phases are either the same or separated by $2\pi$.  Therefore, there is not only a synchronization of spatial motion (coherent breathing), but also a phase synchronization.  The behavior after the early transient can be expressed as $\theta = \Omega t + \delta \theta(t)$, where $\delta \theta(t)$ is an oscillatory function that decays away, leaving behind only the uniform growth of all phases after the transient.  Note that the period of $\delta \theta(t)$ is comparable to $\tau$.  The $\Omega$ can be positive or negative, depending on initial condition.

\begin{figure}[H]
\center \includegraphics[width=4in]{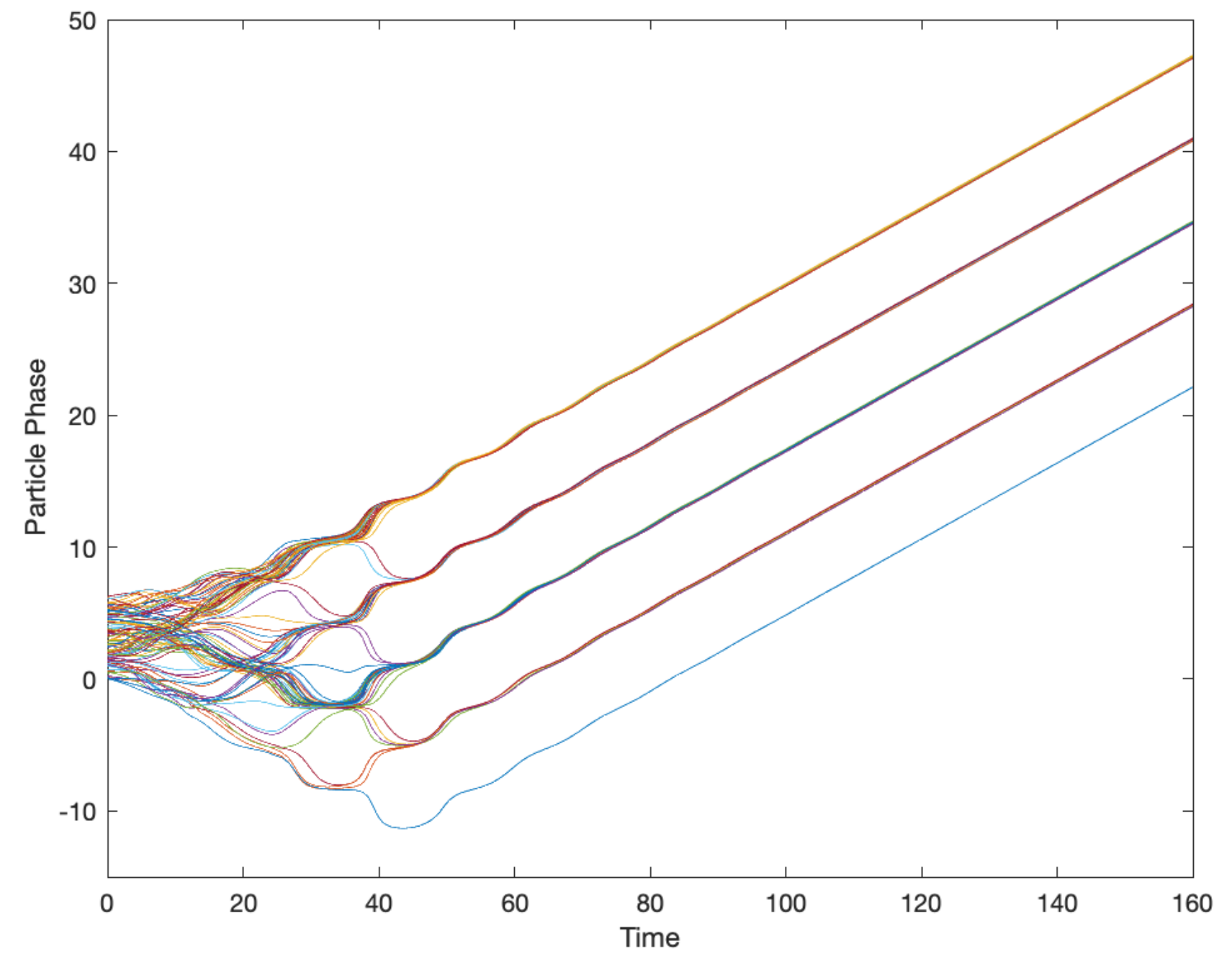}
\caption{$(N,J,K,\tau)=(100,1,-0.75,8)$.  The different groups are separated by $2\pi$}
\label{fig:PhasesOne}
\end{figure}
A detailed example of the time evolution is shown in Fig.~\ref{fig:big_example} in Appendix \ref{sec:Details}.  We also show the evolution of several quantities with increasing $\tau$ in Fig.~\ref{fig:dyn_sim_exs} of the same Appendix, where it is evident that plateaus of $\theta(t)$ become more prominent and grow with increasing $\tau$.

\begin{figure}[H]
\center \includegraphics[width=7in]{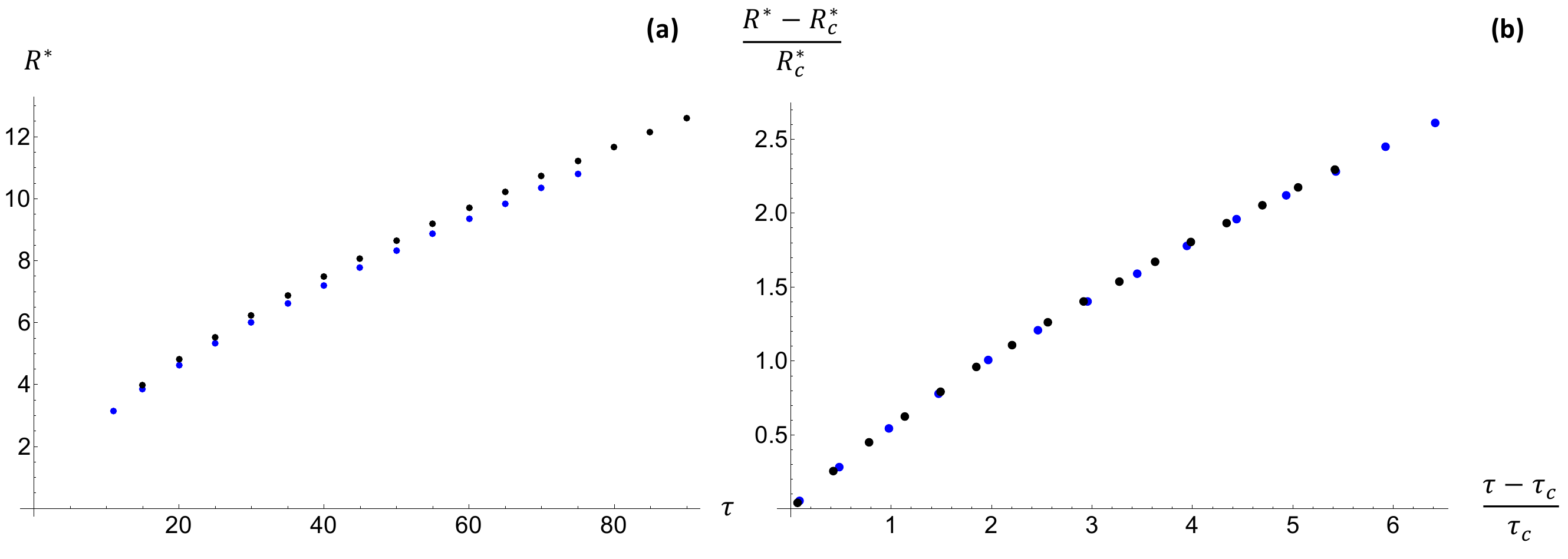}
\caption{(a) $R^*(\tau)$ for two values of $N$.  Top (black) is for $N=1000$ and bottom (blue) is for $N=300$.  The values for $\tau_c$ are $\approx 14$ and $\approx 10.1$, for respective system sizes (see Fig.~\ref{fig:tauc_vs_N}).  Here $J=1$, and $K=-0.7$.  (b) Here the data is plotted versus a rescaled variables.  Evident is the data collapse.  The error bars are not shown because they are very small for this value of parameters.}
\label{fig:R*_vs_tau}
\end{figure}

Before moving on to discuss the boiling collective state, we present data on $R^*(\tau)$ for two different values of $N$.  This is demonstrated in Fig.~\ref{fig:R*_vs_tau}, where $R^*_c$ is the value of $R^*$ at $\tau_c$.  Below $\tau_c$ the value of $R^*$ becomes less well-defined, since it is no longer a static surface, as we will see in the next section.  Note the collapse of the data in Fig.~\ref{fig:R*_vs_tau} (b) unto one universal curve when plotting the dimensionless deviation of the radius $(R^* - R^*_c)/R^*_c$ vs.~ the dimensionless deviation of the delay $(\tau-\tau_c)/\tau_c$.

\subsection{The boiling state}
\label{sec:boiling_num}
\begin{figure}[H]
\center \includegraphics[width=6in]{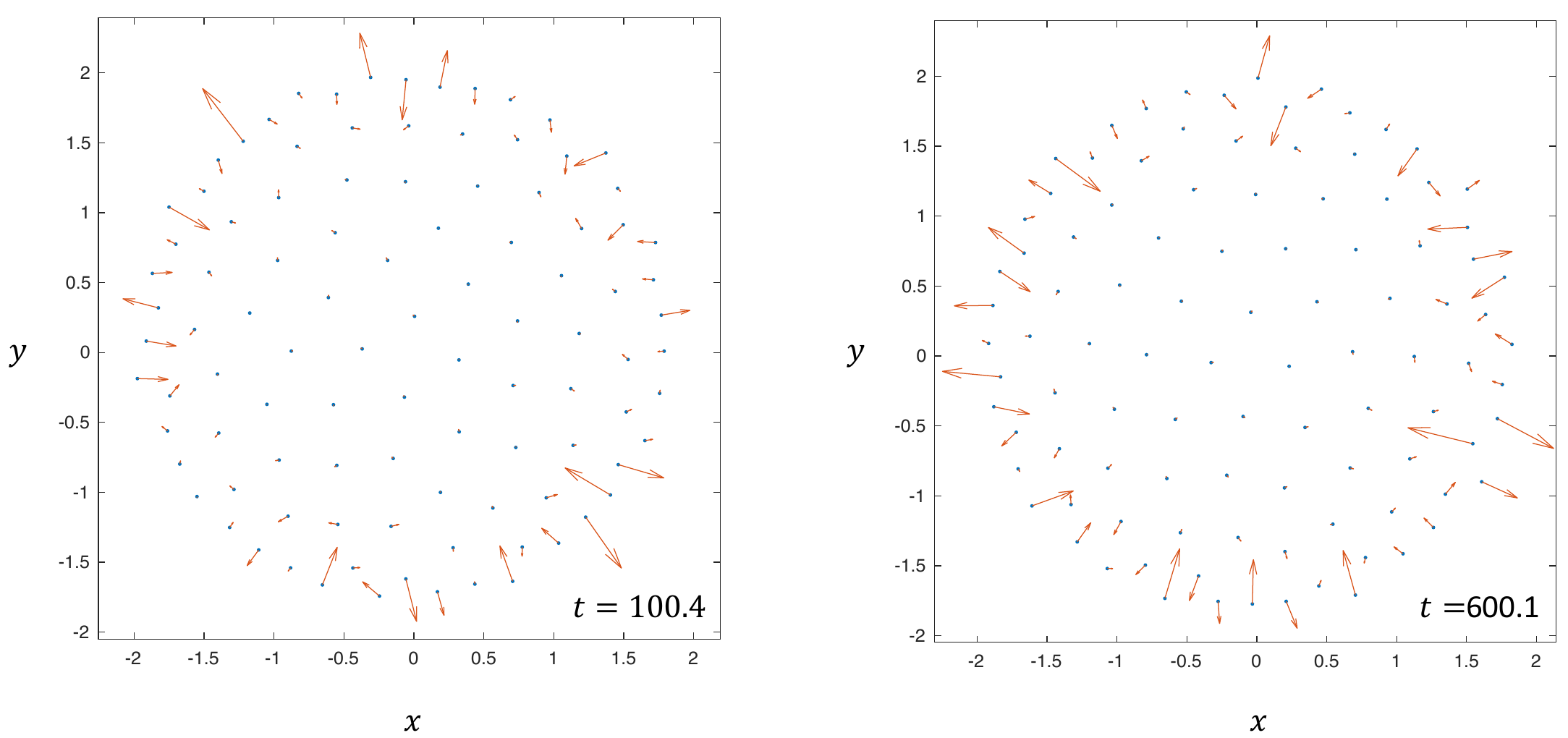}
\caption{Velocity vector plots at two instants of time.  The parameters are $(N,J,K,\tau)=(100,1,-0.75,5)$.}
\label{fig:vectors_boiling}
\end{figure}

At smaller delays, the long breathing part of the transient gives way to a dynamic state, rather than a quasi-static crystal.  In this collective state, swarmalators at the surface of the cluster undergo convective-like motion, while the swarmalators deeper in the interior are essentially frozen, similarly to the pseudo-static situation in the previous subsection.  For this reason, we called it the boiling state, as it looks like the surface of the cluster is boiling.  Fig.~\ref{fig:vectors_boiling} demonstrates several two snapshots of a cluster in such a boiling state.  
We again look at the time evolution of the average radius and average speed.
\begin{figure}[H]
\center \includegraphics[width=7in]{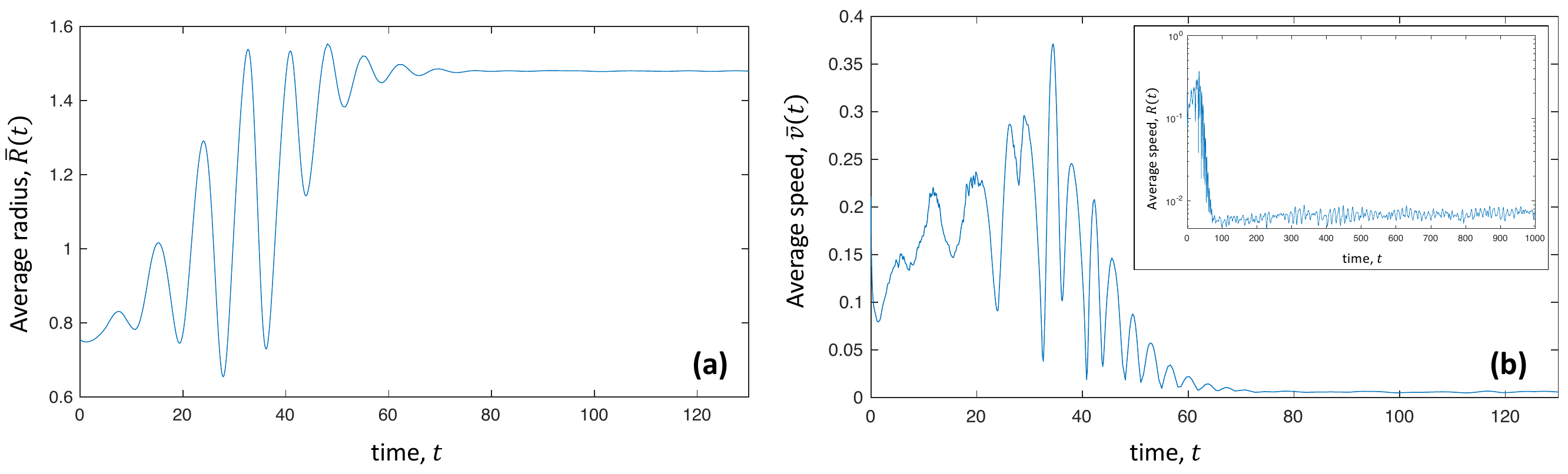}
\caption{(a) $\overline{R}(t)$.  (b) $\overline{v}(t)$.  The parameters are $(N,J,K,\tau )=(100,1,-0.75,5)$}
\label{fig:avg_R_and_avg_v_boiling}
\end{figure}
In contrast to the situation at larger $\tau$, there is a residual average velocity.  It is not the same as the type of residual motion that was left after the breathing transient at larger $\tau$.  The type of motion in the boiling state is qualitatively different, and the value of the average velocity is also larger by at least an order of magnitude.

The dynamics of internal phases is similar to the the phase dynamics in the higher $\tau$ collective state.
\begin{figure}[H]
\center \includegraphics[width=4in]{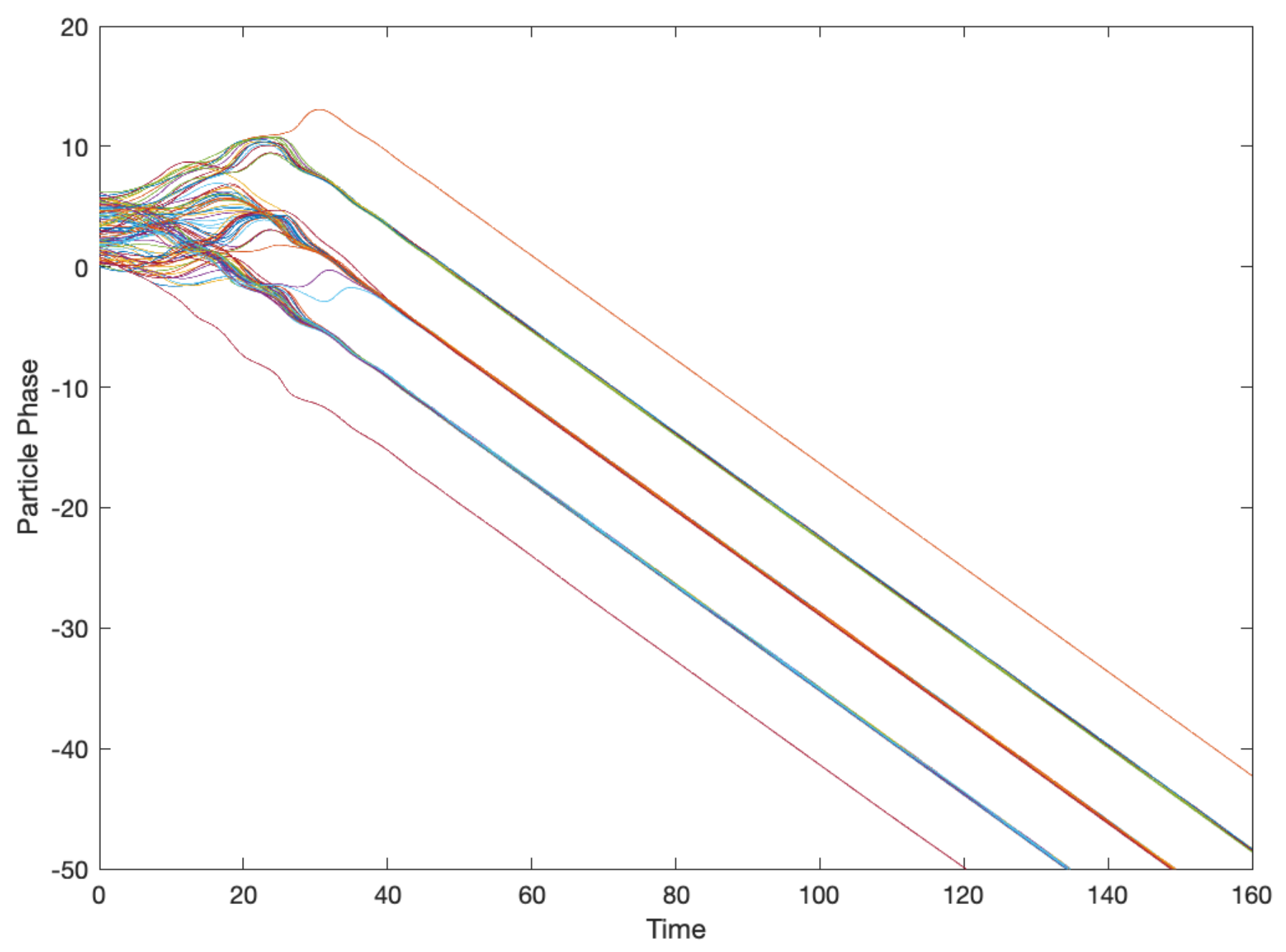}
\caption{$(N,J,K,\tau)=(100,1,-0.75,5)$.   The different groups are separated by $2\pi$.}
\label{fig:PhasesTwo}
\end{figure}
%

\subsection{Delay-induced transition}
\label{sec:transitions}
\begin{figure}[H]
\center \includegraphics[width=4in]{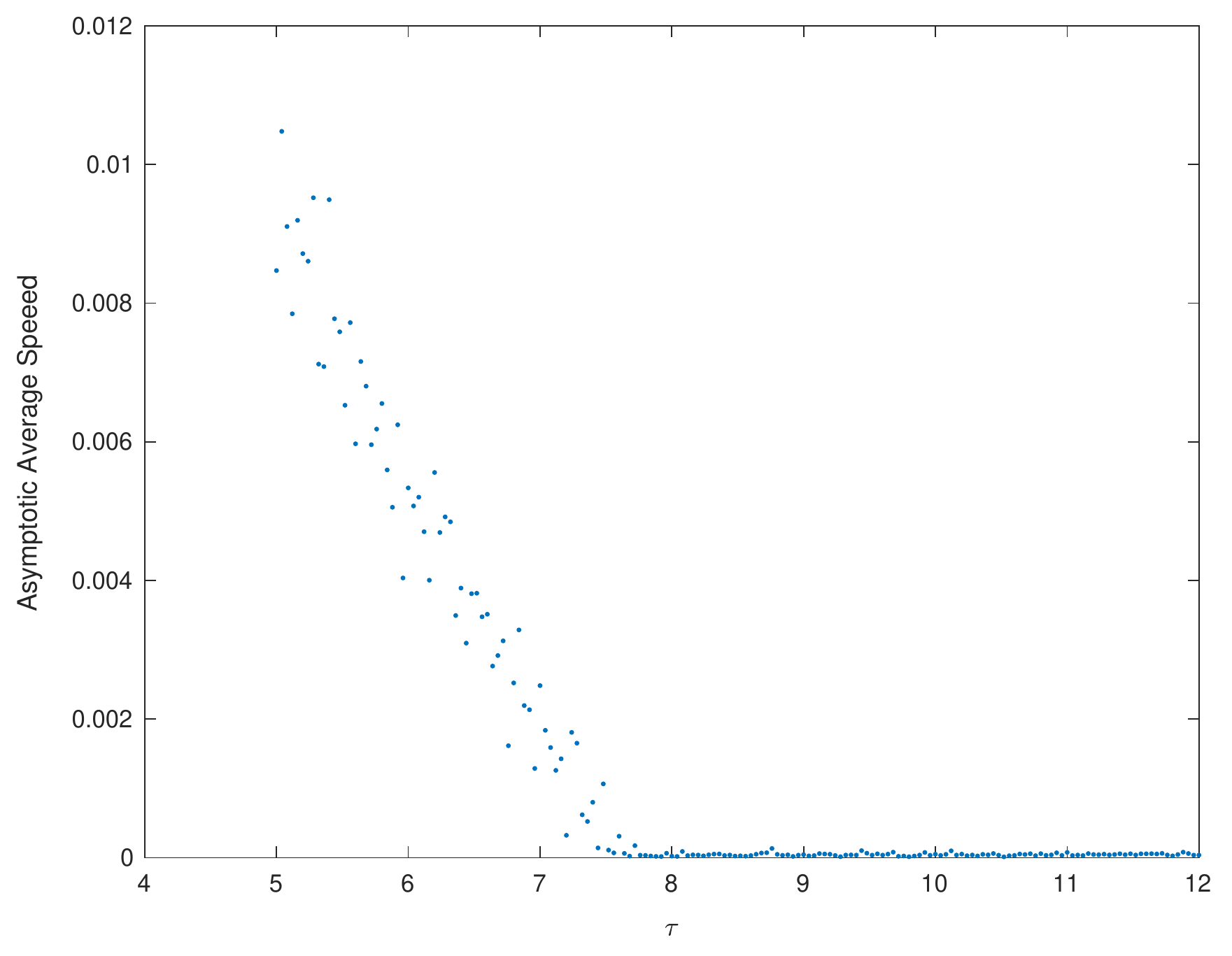}
\caption{$(N,J,K)=(100,1,-0.7)$.  The data was collected at $t=1500$ after the initial condition.}
\label{fig:transition_intro}
\end{figure}

\begin{figure}[H]
\center \includegraphics[width=4in]{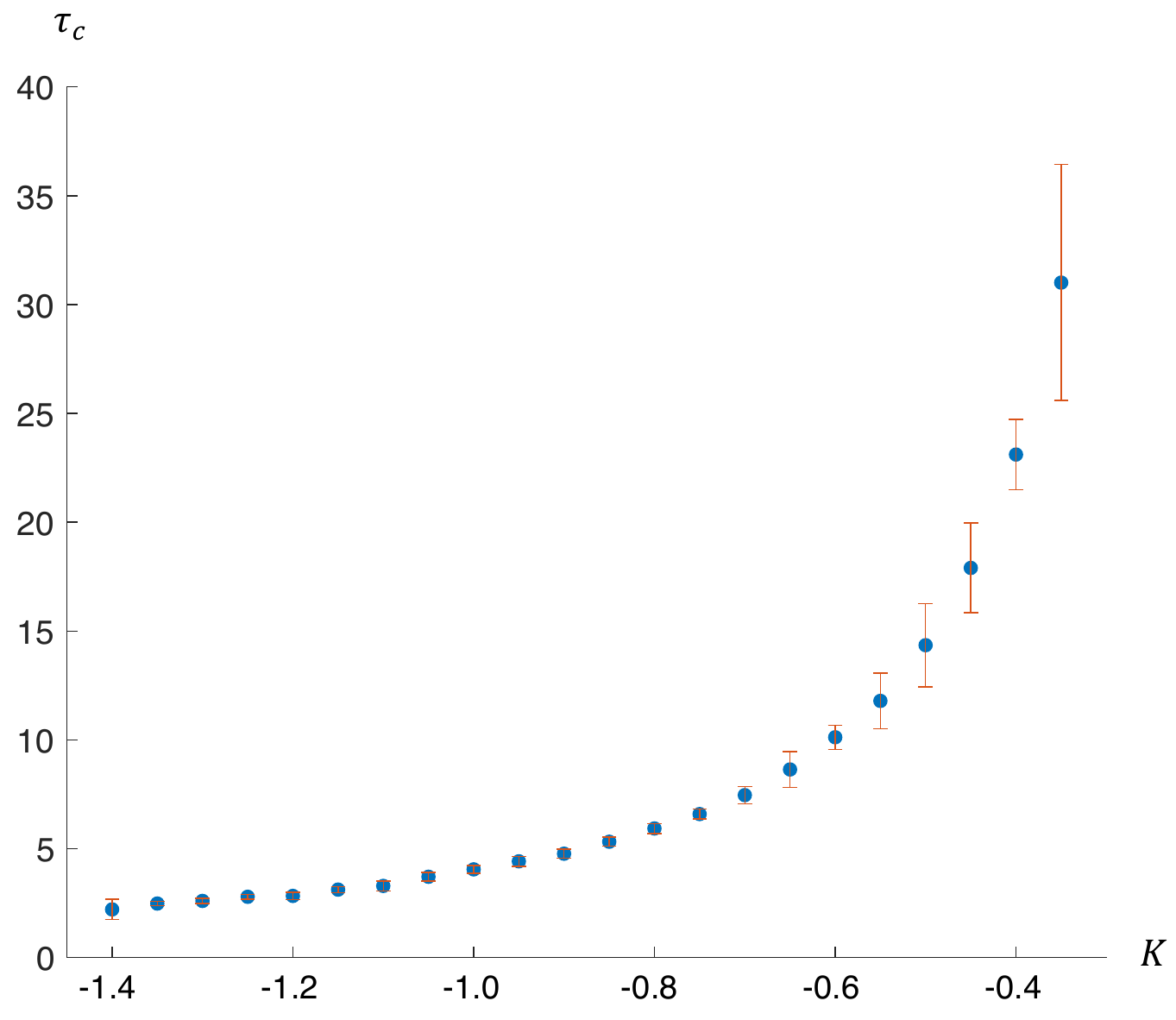}
\caption{$(N,J)=(100,1)$.}
\label{fig:tauc_vs_K}
\end{figure}

We presented two types of collective states that develop at large times - the pseudo-crystaline phase at larger $\tau$ and the boiling state at smaller $\tau$.  We will now discuss the transition between these states.  Consider the plot of the the large time asymptotic average speed versus the delay time $\tau$, see Fig.~\ref{fig:transition_intro}.
There is a well-defined transition at a certain $\tau$.  It corresponds to the transition between the boiling state at smaller $\tau$ and the quasi-static state at larger $\tau$.  The fluctuations seen in the values of the asymptotic speed result from different initial conditions from one simulation to the next.  Even if particle positions and internal phases are identical at all $\tau$, changing the value of $\tau$ itself will lead to completely different trajectories $\textbf{x}_i(t)$ and $\theta_i(t)$ between even very nearby $\tau$ (similar to, or perhaps identical to chaotic divergence).  The critical $\tau$ can be extracted by fitting the lower-$\tau$ portion of the graph by a straight line and extracting the $\tau$ of the $x$-intercept. The following  figure presents $\tau_c$ thus extracted from numerical calculations versus the coupling strength $K$ at $J=1$ - see Fig.~\ref{fig:tauc_vs_K}.
There appears to be a minimal value of $|K|$ below which it is impossible to induce breathing behavior no matter the value of $\tau$.  As $|K|$ decreases, and  approaches this minimal $|K|$, the fluctuations grow.  Both of these observations remind us of critical phenomena. 

\begin{figure}[H]
\center \includegraphics[width=7in]{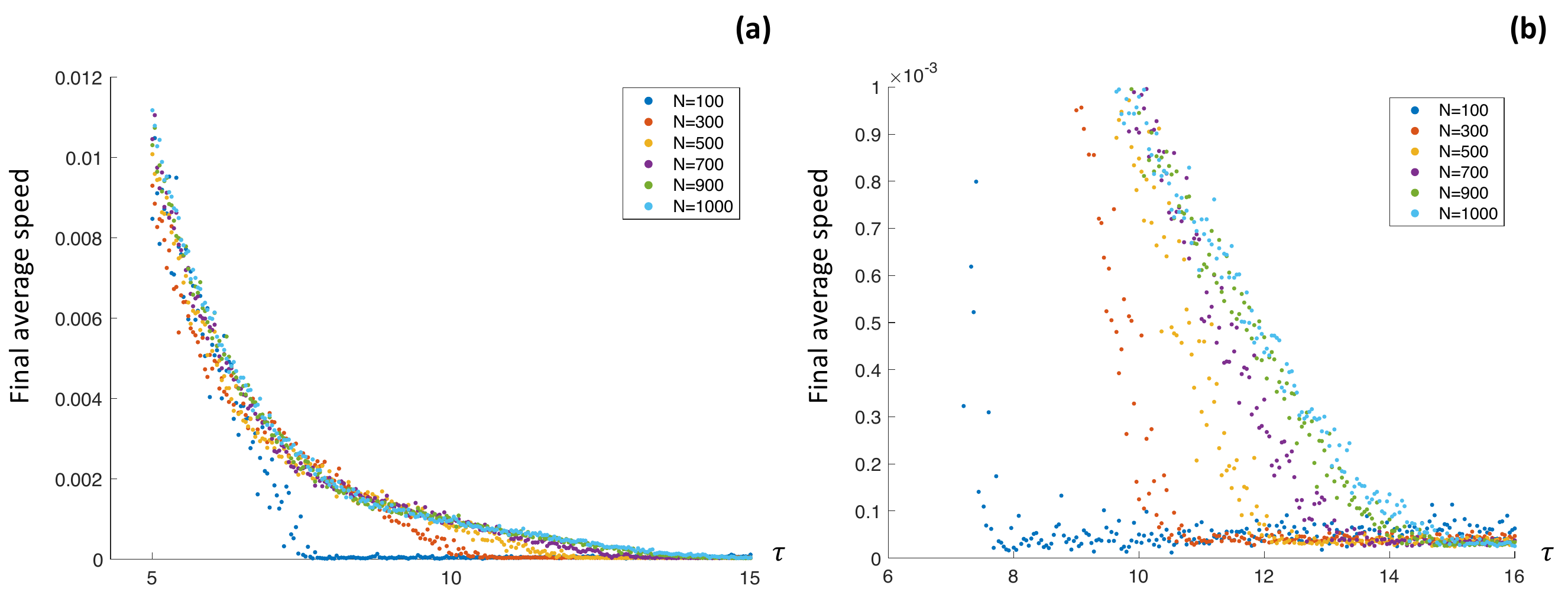}
\caption{(a) Final average speed vs.~$\tau$ for several $N$.  The parameter values are$(J,K)=(1,-0.7)$.  The data was collected at $t=1500$ after the initial condition.  (b)The data was collected at $t=1500$ after the initial condition.  The values do not reach all the way to zero because of the residual velocities after the breathing, as described above (note that the lowest values are $\sim 5 \times 10^{-5}$, which is comparable to what we see in Fig.~\ref{fig:residual}).}
\label{fig:transition_multipleN}
\end{figure}

\begin{figure}[H]
\center \includegraphics[width=4in]{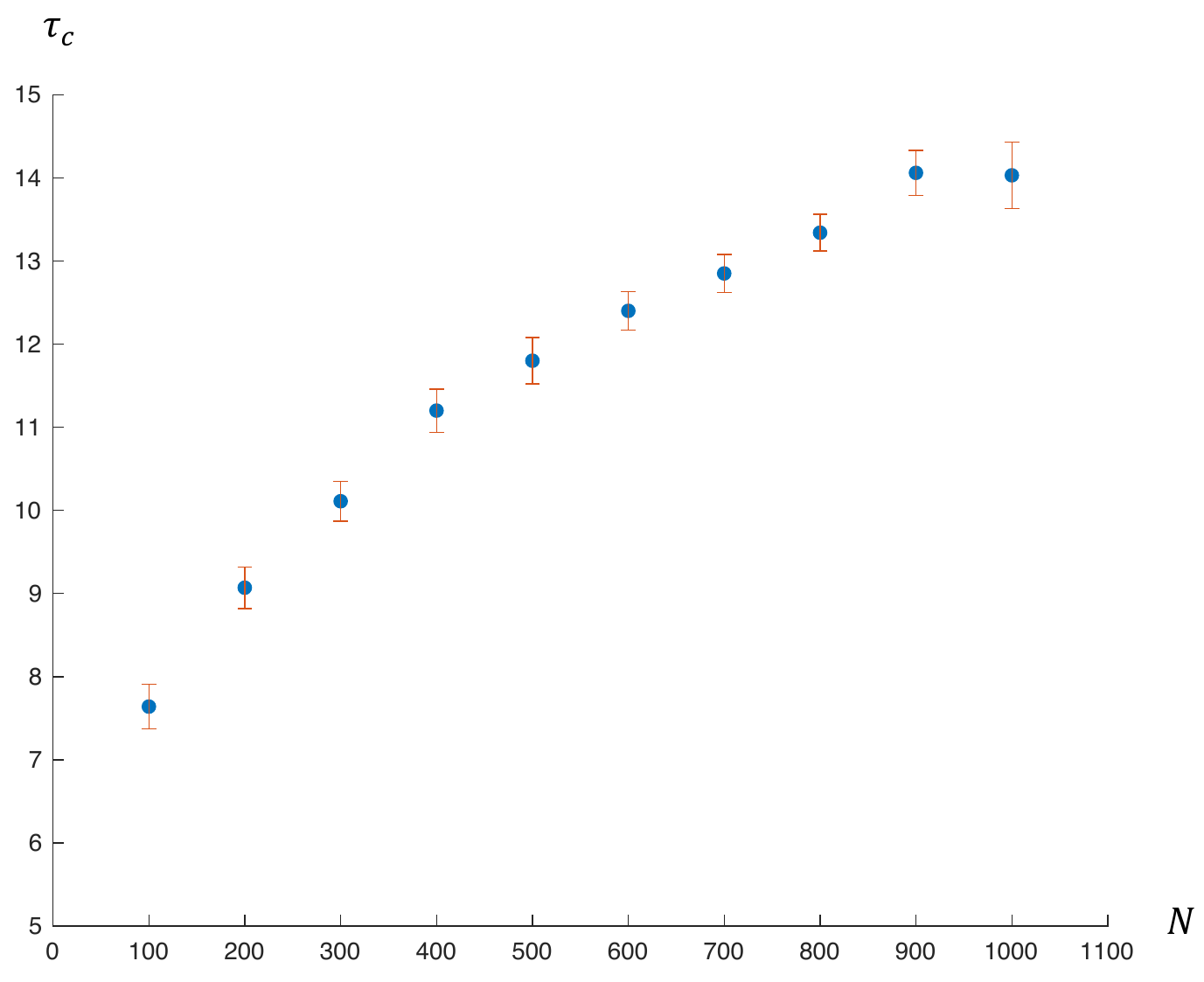}
\caption{$\tau_c(N)$.  Here $(J,K)=(1,-0.70)$. The data was collected at $t=1500$ after the initial condition.}
\label{fig:tauc_vs_N}
\end{figure}

We will now demonstrate how the transition from breathing to boiling depends on the system size.  First, Fig.~\ref{fig:transition_multipleN} is analogous to Fig.~\ref{fig:transition_intro}, but includes data on progressively larger system sizes.  If we restrict the range of the $y$-axis, it becomes clear that $\tau_c$ displays a trend towards a limiting value.
%
This saturation is also evident in Fig.~\ref{fig:tauc_vs_N}
It is important to emphasize that oscillations persist below $\tau_c$.  Above $\tau_c$, oscillations give way to the quasi-static pseudo-crystal at large times.  Below $\tau_c$, oscillations give way to the boiling state at large times.  Thus, $\tau_c$ delimits two large-time (or infinite time) behaviors after the breathing transient has subsided;  it does not refer to properties of breathing oscillations.  Fig.~\ref{fig:moustail_evolution} in Appendix \ref{sec:Details} demonstrates the evolution of the average speed and the average radius across $\tau$, from above $\tau_c$ to below $\tau_c$.  We see that as $\tau$ crosses below $\tau_c$, the boiling layer begins to develop.  As $\tau$ is progressively decreased, the thickness of the boiling layer grows.  The long-time value of the average radius becomes less and less well defined, since the boiling of the surface leads to increasing fluctuations of this quantity.  However, as long as the system is in the boiling regime, the phases synchronize.  At even lower values of $\tau$, we would encounter another transition, $\tau_l$, below which the phases no longer synchronize.  In the region of $(J,K)$ parameter space in which we have done numerical investigations, we found that in this lower $\tau$ regime the dynamical behavior resembles active phase waves.  However, the situation in other  regions of parameter space may be different; most of our numerical exploration took place in the $(J,K)$ region that corresponds to the active phase waves in the absence of the delay.

Thus, the frequency $\omega$ and decay rates $r$ of breathing oscillations should smoothly vary across $\tau_c$;  $\tau_c$ refers to the large-time behavior, not to properties of breathing oscillations.  We now present the numerical results for $\omega$ and $r$.
\begin{figure}[H]
\center \includegraphics[width=5in]{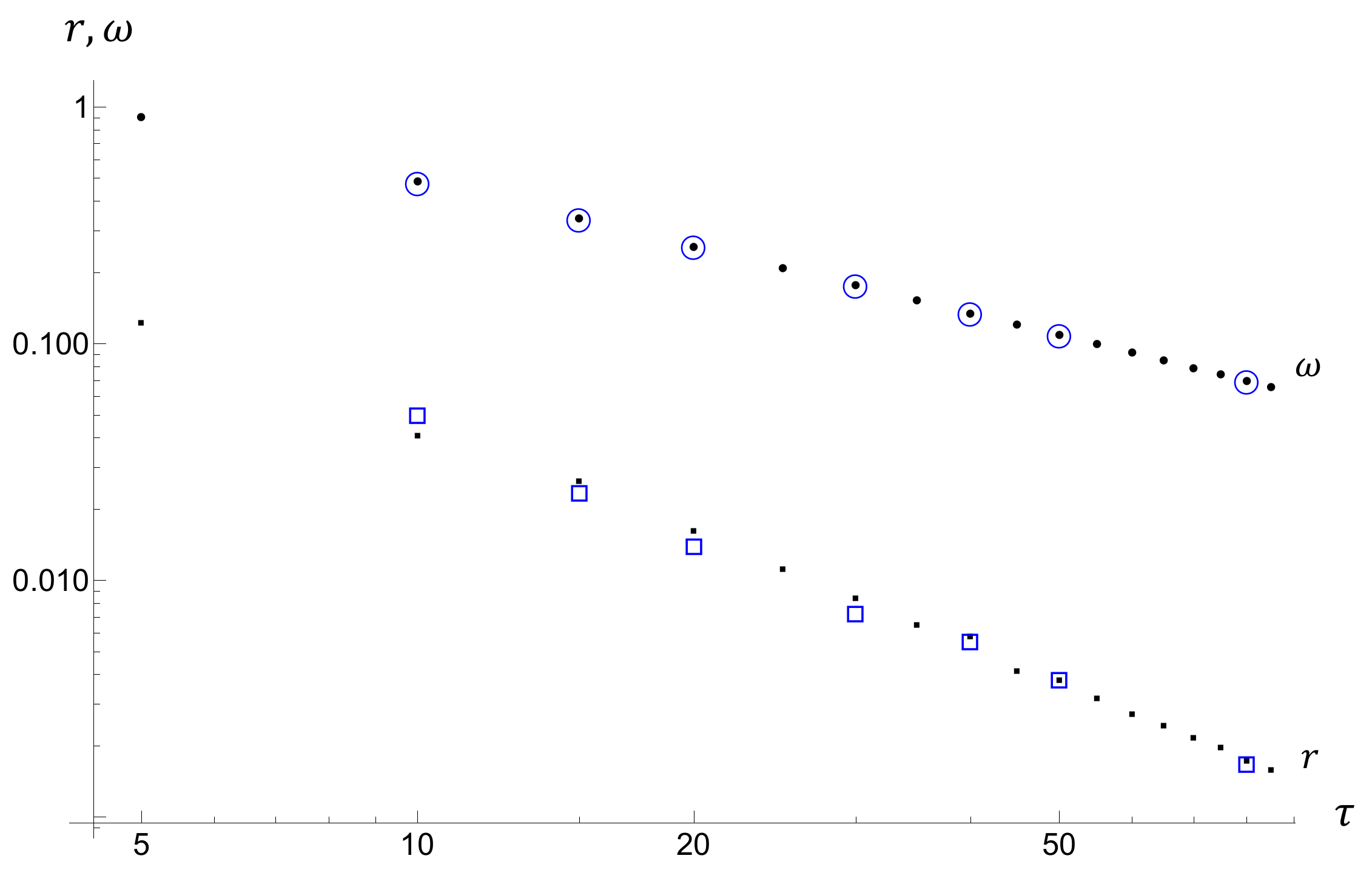}
\caption{Decay rates $r$ (squares) and angular frequencies $\omega$ (circles) of breathing oscillations obtained from numerical simulations.  Blue open symbols, $N=300$.  Black solid symbols, $N=1000$.}
\label{fig:rates_omegas_numerics}
\end{figure}
We make two important remarks.  First, note that in contrast to Fig.~\ref{fig:R*_vs_tau}(a), the data for $N=300$ and $N=1000$ follows  essentially the same functional dependence, so there is no need to rescale the variables to achieve data collapse; there is already a data collapse as is.  Second, we see that this collapse is better for $\omega$ than for $r$.  One might also ask why the data was not collected at lower $\tau$s.  In order to extract this pair of parameters (for example, $\omega$ and $r$), we have to fit the late time tail of $R(t)$ data generated by the simulation to the functional form $Ae^{-rt}\cos{(\omega t + \phi)} + B$. Or, we have subtract off the $B$ from the data first and then make the fit with zero $B$.  In either case, we have to know the $B$ from the long-time asymptote towards which $R(t)$ relaxes.  However sufficiently below $\tau_c$, that asymptote is noisy due to the fluctuations of the surfaces, as explained above.  At the same time, the decay rate grows.  Thus, these parameters estimated from such a fit become less and less accurate at lower $\tau$.  Moreover, at a certain $\tau$ there is simply not enough of data on which a fit can be made.  We will see below that in the theoretical approach there is no plague of fluctuations, and so the prediction for $r(\tau)$ and $\omega(\tau)$ can be made for lower values of $\tau$.

\newpage
\section{Theoretical understanding of collective breathing}
\label{sec:Theory}
Our simulations revealed for $\tau$ on both sides of $\tau_c$ the phase slips decay away at long times, and all phases advance uniformly with rate $\Omega$.  We also observed that this is accompanied by freezing out of spatial motion, i.e. all $\vec{r}_i$ reach a constant value $\vec{r}^*_i$.  Therefore, at large times, Eq.~(\ref{eq:theta_eq}) becomes 
\begin{equation}
\label{eq:single_sum2}\Omega = -\sin{(\Omega \tau)} \frac{K}{N} \sum_{j \neq i} \frac{1}{|\vec{r}^*_j - \vec{r}^*_i|} 
\end{equation}
The sum is some constant number.  On the one hand, it seems to depend on the position of this $i$-th oscillator.  On the other hand, it equals a constant that's independent of $i$, so this will translate to a self-consistency argument on the density of oscillators, which we will analyze below.

\begin{figure}[H]
\center \includegraphics[width=3in]{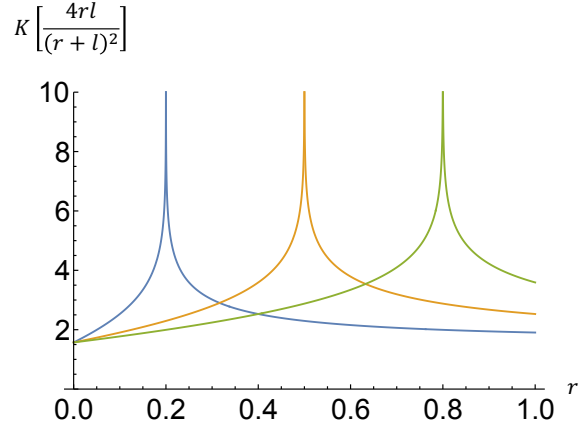}
\label{fig:Kernel_function}
\caption{$K\left[\frac{4lr}{(r+l)^2}\right]$ as a function of $r$ for $l=0.2$ (left), $0.5$ (middle) and $0.8$ (right).}
\end{figure}

To calculate the sum, we will define $\rho(r)$ to be the equilibrium density of swarmalators, such that $\rho(r)r\,dr\,d\theta$ is the number of swarmalators in a differential area $r\,dr\,d\theta$.  We have assumed a radial symmetry, which invites the use of polar variables, and is the reason why $\rho$ is a function of only the radius.  This assumption conforms to our observations and it is expected because there are no symmetry-breaking fields.  Consider two swarmalators: swarmalator $i$, located at radius $l$, and swarmalator $j$, located at radius $r$ and angle $\alpha$.  Because of radial symmetry, the sum will not depend on the angle of swarmalator $i$, so it is convenient to place it at zero angle relative to an arbitrary $x$-axis.  The situation is depicted in the following schematic.  
\begin{figure}[H]
\center \includegraphics[width=3in]{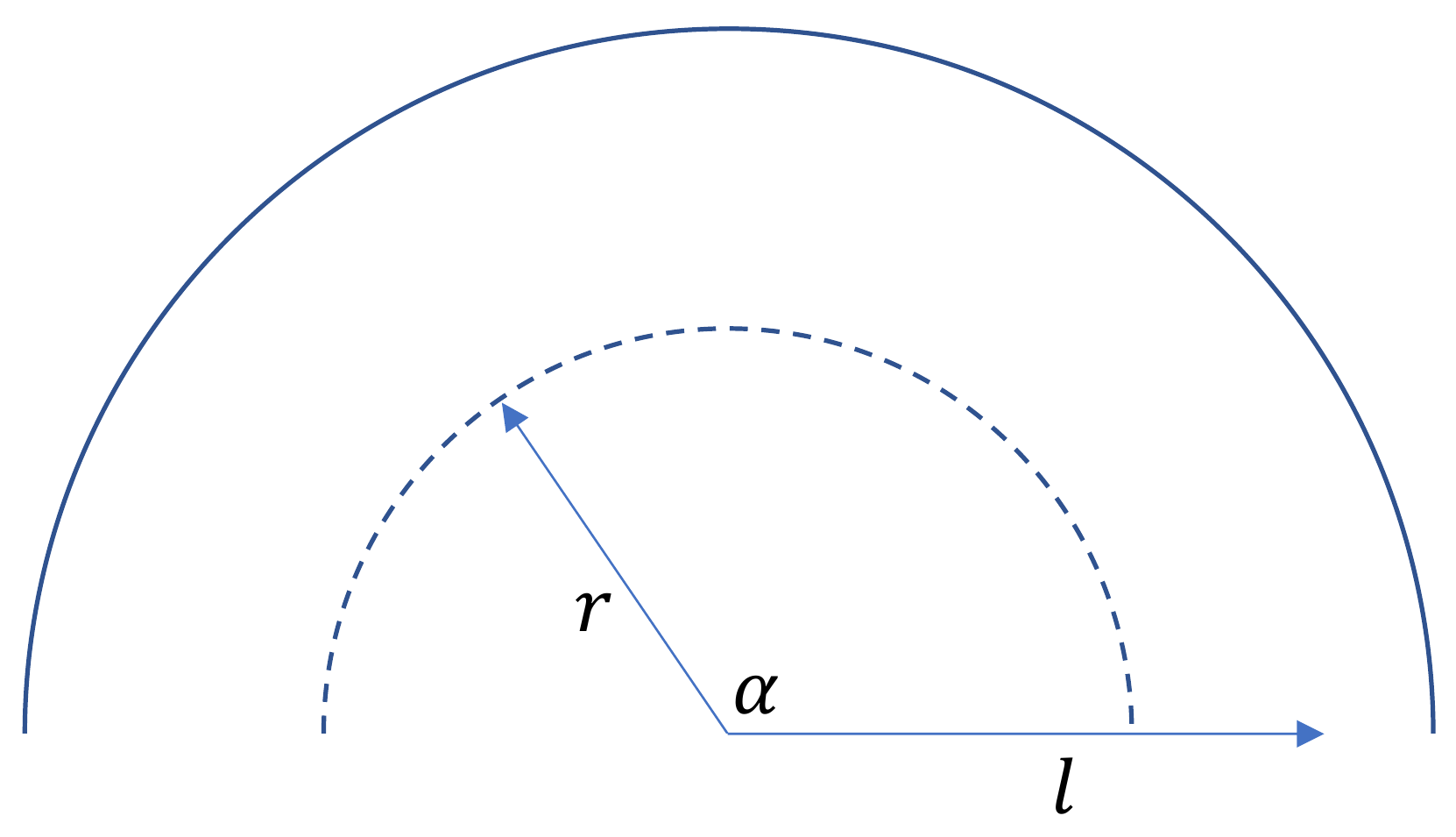}
\caption{Geometry for evaluating Eq.~(\ref{eq:single_sum2}).}
\label{fig:six}
\end{figure}
With this preparation, the sum in Eq.~(\ref{eq:single_sum2}) (we will call it $s$) can be calculated with the help of the density $\rho(r)$ in the following way
\begin{equation}
s =  2\int_0^R dr \int_{0}^{\pi} \frac{\rho(r)r d\alpha}{\sqrt{r^2 + l^2 -2r l\cos{\alpha}}}
\end{equation}
Here $R$ is the radius of the whole cluster.  The integral over $\alpha$, which serves as the kernel for the integral over $r$, evaluates to $\frac{2K\left[\frac{4lr}{(r+l)^2}\right]}{(r+l)}$, where $K$ is a complete elliptic integral of the first kind.  It diverges when its argument is $1$ - which will happen when $r=l$.  We depict $K\left[\frac{4lr}{(r+l)^2}\right]$ vs.~$r$ for several $l$ in Fig.~\ref{fig:Kernel_function}.
To continue further, we need to know $\rho(r)$, which is not given a-priori.  However, we can determine it self-consistently.  Note that integrals are functions of $l$, the radius of $i$-th swarmalator.  On the other hand, the left hand side of Eq.~(\ref{eq:single_sum2}) must be independent of $l$ - it must be the same for all swarmalators.  Therefore, $\rho(r)$ must be a special function that will ensure that the answer is independent on $l$.  Before seeking this self-consitent solution, it helps to reason physically what we expect such a special $\rho(r)$ to be.  Eq.~(\ref{eq:single_sum2}) is a sum of $1/distance$ between swarmalator $i$ and all the other swarmalators $j$.  In the vicinity of the center of the swarm, this sum is essentially invariant as we sample different $i$ around this center.  Closer to the edge, essentially a large part of the sum is missing - about a half of the swarm is missing.  So, to compensate for this $\rho$ must increase towards the edge if we are to have the same value of the sum for all $i$s.

\begin{figure}[H]
\center \includegraphics[width=4.5in]{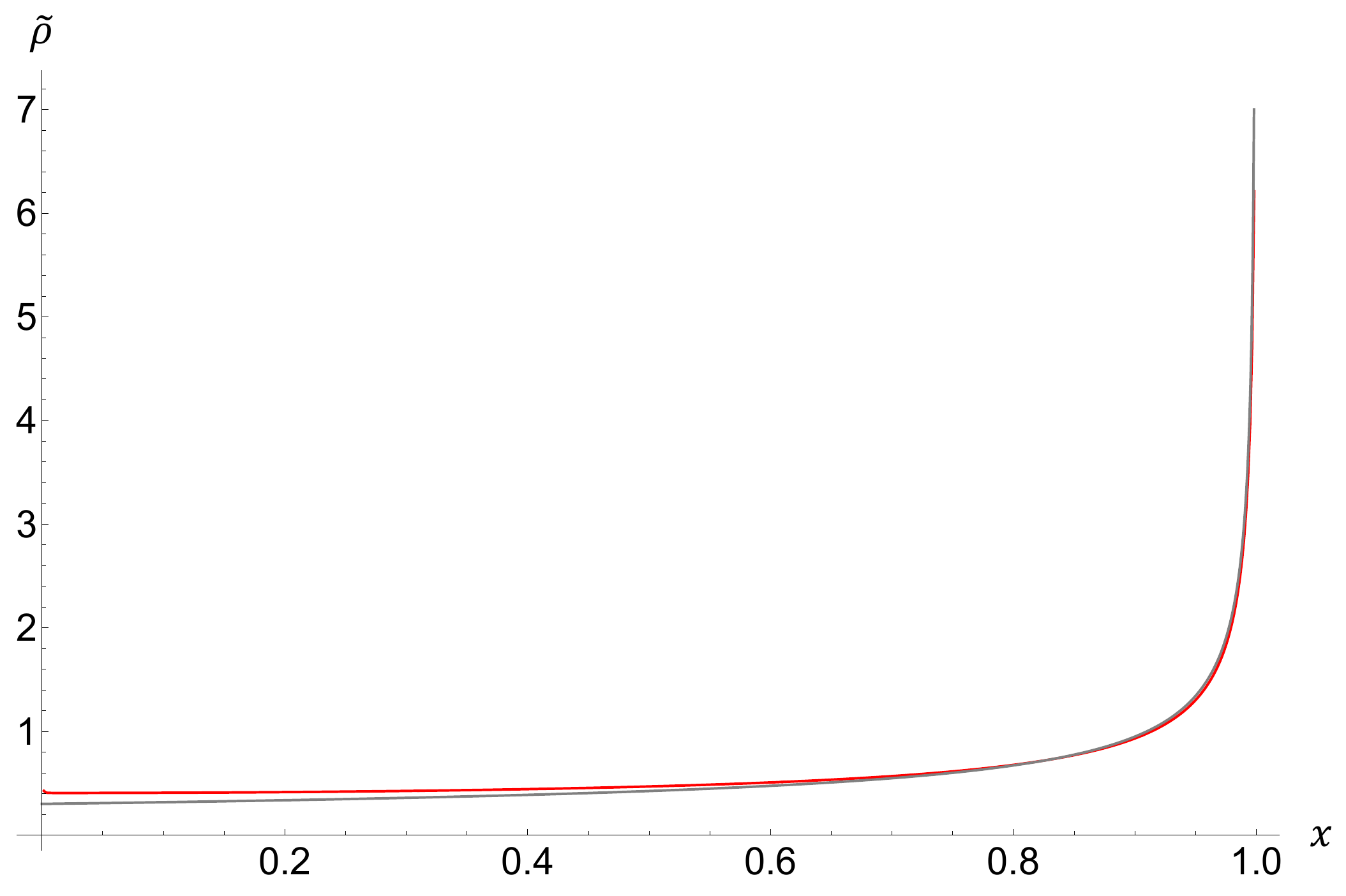}
\caption{Red (slightly higher curve at small $x$): $\tilde{\rho}$ obtained from the solution of Eq.~(\ref{eq:matrix}), Gray: $0.3 (1-x)^{-1/2}$.}
\label{fig:Rho_plot}
\end{figure}

With we now seek $\rho(r)$ self-consistently.  At this point in the calculation, Eq.~(\ref{eq:single_sum2}) has become
\begin{equation}
\label{eq:transformed1} -\frac{N\Omega}{4K\sin{(\Omega \tau)}}  = \int_0^R r\rho(r) \frac{K\left[\frac{4lr}{(r+l)^2}\right]}{(r+l)} \,dr.
\end{equation}
We can clean it up by letting $\rho = -\frac{N\Omega}{4RK\sin{(\Omega \tau)}} \tilde{\rho}(r)$, and also changing variables $r=Rx$ and $l = RL$.  Eq.~(\ref{eq:transformed1}) becomes 
\begin{equation}
\label{eq:rho-tilde_equation}
1 =  \int_0^1 x\tilde{\rho}(x) \frac{K\left[\frac{4Lx}{(x+L)^2}\right]}{(x+L)} \,dx = \int_0^1 \tilde{\rho}(x) \mathcal{K}(L,x)\,dx
\end{equation}
Our objective is to find such $\tilde{\rho}(x)$ that makes this true for any $L$.  We do this numerically, by approximating the integral by a discrete sum, giving the following set of equations:
\begin{eqnarray}
1 &=& \left(\tilde{\rho}(x_1)\mathcal{K}(L_1,x_1) + \tilde{\rho}(x_2)\mathcal{K}(L_1,x_2) + ... \tilde{\rho}(x_n)\mathcal{K}(L_1,x_n)\right)\Delta x \nonumber \\
1 &=& \left(\tilde{\rho}(x_1)\mathcal{K}(L_2,x_1) + \tilde{\rho}(x_2)\mathcal{K}(L_2,x_2) + ... \tilde{\rho}(x_n)\mathcal{K}(L_2,x_n)\right)\Delta x \nonumber \\ 
&...& \nonumber \\
1 &=& \left(\tilde{\rho}(x_1)\mathcal{K}(L_n,x_1) + \tilde{\rho}(x_2)\mathcal{K}(L_n,x_2) + ... \tilde{\rho}(x_n)\mathcal{K}(L_n,x_n)\right) \Delta x \label{eq:matrix}
\end{eqnarray}
We construct a matrix of kernels $\mathcal{K}(L_n,x_m) \Delta x$, then invert this matrix, operate on the vector $(1,1, ... , 1)$, and find a vector $(\tilde{\rho}(x_1), \tilde{\rho}(x_2), ... , \tilde{\rho}(x_{n_{max}}))$.   Because $\mathcal{K}$ diverges when its argument in $1$, the same values could not be used for $L$s and $x$s.  We choose $x_n = \frac{n}{n_{max}}$ and $L_n = \frac{n}{n_{max}} + 10^{-4}$, where $n_{max}$ was typically on the order of a thousand.  The result of this calculation with $n_{max} = 600$ is displayed in Fig.~\ref{fig:Rho_plot} in red.  

The density increases at the edge, as expected.  The divergence at $x=1$ appears to have an exponent very close to $1/2$, i.e. to have the form $(1-x)^{-1/2}$ close to $x=1$.  The function $\tilde{\rho}$ is not a pure power law over the entire domain $[0,1]$.  However, we would like to model this result by an analytical expression in order to make tractable, analytical predictions.  We found that the function $0.3(1-x)^{-1/2}$ approximates the whole numerical result $\tilde{\rho}(x)$ very well - see Fig.~\ref{fig:Rho_plot}.
We checked that $0.3 (1-x)^{-1/2}$ makes the integral $\int_0^1 \rho(x) \mathcal{K}(L,x) \,dx$ very nearly a constant with respect to $L$ (keeping in mind that $0.3(1-x)^{-1/2}$ is an approximation).

The total number of swarmalators inside the cluster should be equal to $N$.  Therefore
\begin{eqnarray*}
N = 2\pi \int_0^R \rho(r) r \,dr &=& -2\pi \frac{N\Omega}{4RK\sin{(\Omega \tau)}} \int_0^R r \tilde{\rho}(r) \,dr \\
&=& - \frac{N \pi \Omega}{2RK\sin{(\Omega \tau)}} R^2 \int_0^1 x \tilde{\rho}(x) \,dx \\
&\approx& - \frac{NR \pi \Omega}{2K\sin{(\Omega \tau)}} \int_0^1 0.3x (1-x)^{-1/2} \,dx \\
&=& -\frac{NR \pi \Omega}{5K\sin{(\Omega \tau)}}
\end{eqnarray*}
(the integral evaluates to $0.4$).  So,
\begin{equation}
\label{eq:R-Omega_1}
\frac{\Omega}{\sin(\Omega \tau)} = -\frac{5K}{R\pi}
\end{equation}
Therefore, 
\begin{equation}
\rho(r) = \frac{5}{4}\frac{N}{\pi R^2} \tilde{\rho}(r) 
\label{eq:Rho_Model} 
\end{equation}
This $\rho(r)$ is normalized, i.e. $2\pi \int_0^R r \rho(r) \,dr = N$.  The expression for $\rho$ in Eq.~(\ref{eq:Rho_Model}) is not complete, because we don't know how $R$ depends on $N$ and other parameters.  So, we need more information in order to close this expression and make it self-contained.  We will attempt to use the spatial equation for this purpose.  
%
The $\tilde{\rho}$ is a parameter-free, dimensionless function that's a solution to Eq.~(\ref{eq:rho-tilde_equation}).  We obtained it numerically by discretizing the radial variable and insisting that the integral in Eq.~(\ref{eq:rho-tilde_equation}) should be independent of $L$, and always gives $1$.

The radial density function is $\mathcal{R}(r) = 2\pi r \rho(r) = \frac{10}{4}\frac{N}{R^2}r \tilde{\rho}(r)$.  This is expressed as a function of $r$ that goes between $0$ and $R$.  If we want to express it as a function of a dimensionless variable $x\equiv \frac{r}{R}$, that goes between $0$ and $1$, the answer is $\mathcal{R}(x) = \frac{10}{4}Nx\tilde{\rho}(x)$.  
It might at first seem that a factor $R$ is missing from the denominator, but in fact it is this form that normalizes to $N$.  In other words,
\begin{eqnarray*}
\int_0^1 \mathcal{R}(x) \,dx &=& \frac{10}{4}N \int_r^R \frac{r}{R} \tilde{\rho}(r) \frac{dr}{R} \\
&=& \frac{10}{4} \frac{N}{R^2} \int_0^R r\tilde{\rho}(r) \,dr \\
&=& \frac{5}{4}\frac{N}{\pi R^2} \int_0^R 2\pi r\tilde{\rho}(r) \,dr \\
&=& \int_0^R 2\pi r \rho(r) \,dr \\
&=& N.
\end{eqnarray*}
Note that $\frac{dr}{R}$ in the integration measure is what required the correct definition of $\mathcal{R}(x)$ to not have the $R$ in the denominator.  In doing the analysis of data, we also normalize the histogram so that $\sum_i{\rho_i}\Delta x = N$. 

\begin{figure}[H]
\center \includegraphics[width=6in]{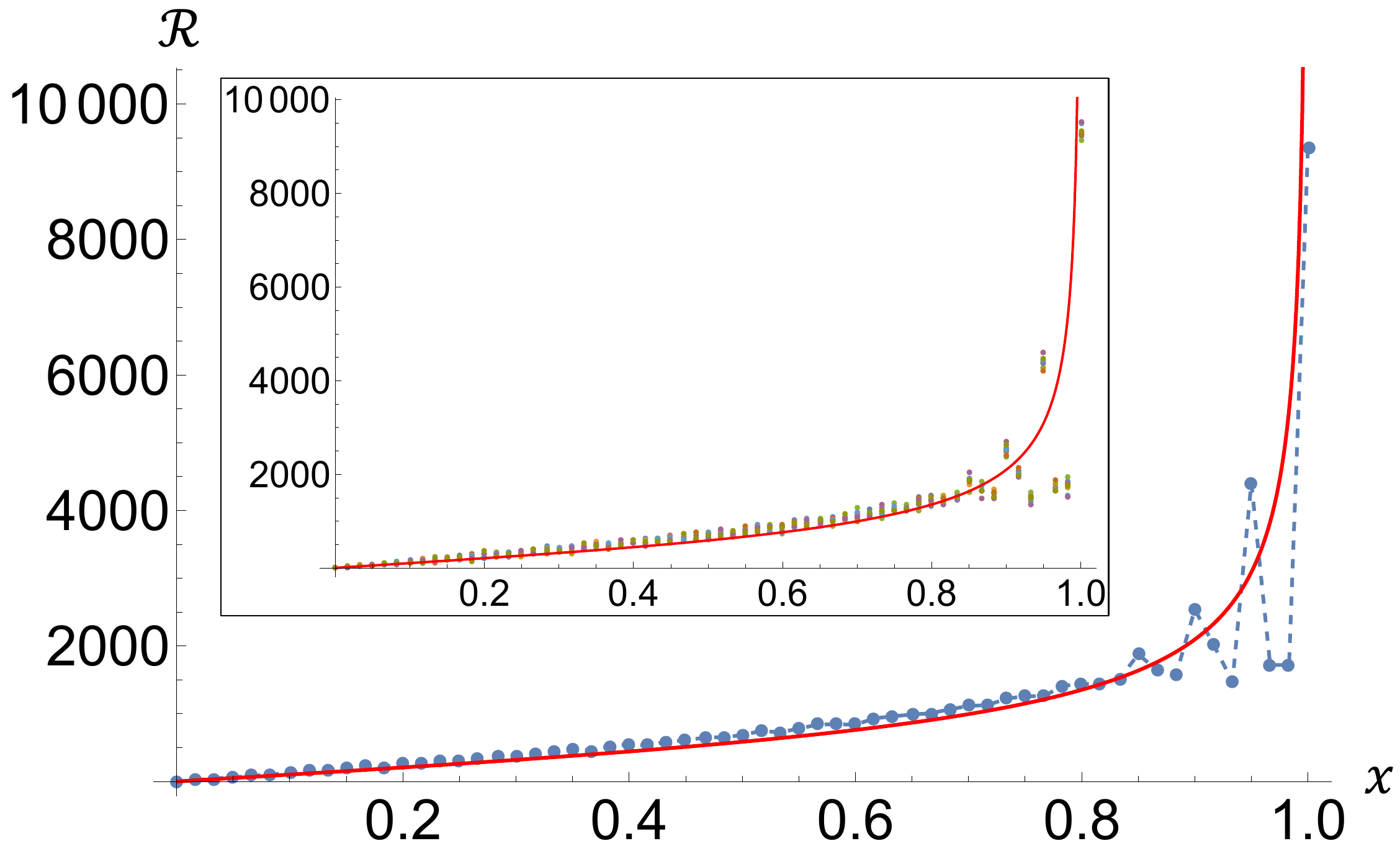}
\caption{Inset: there are ten sets of dots, each with a unique color.  Each set represents the average over ten simulations, and corresponds to one unique value of $\tau$ from $17.0$ to $17.9$.  Each set of dots represents the number of particles within each radial bin.  There are $61$ bins.  The solid curve is $\mathcal{R}(x)$ predicted by theory.  The main plot contains only one set of dots, which is the average of all ten.  The dots are connected by a dashed line to guide the eye to oscillations near $x=1$.}
\label{fig:Densities_comparison1}
\end{figure}

\begin{figure}[H]
\center \includegraphics[width=7in]{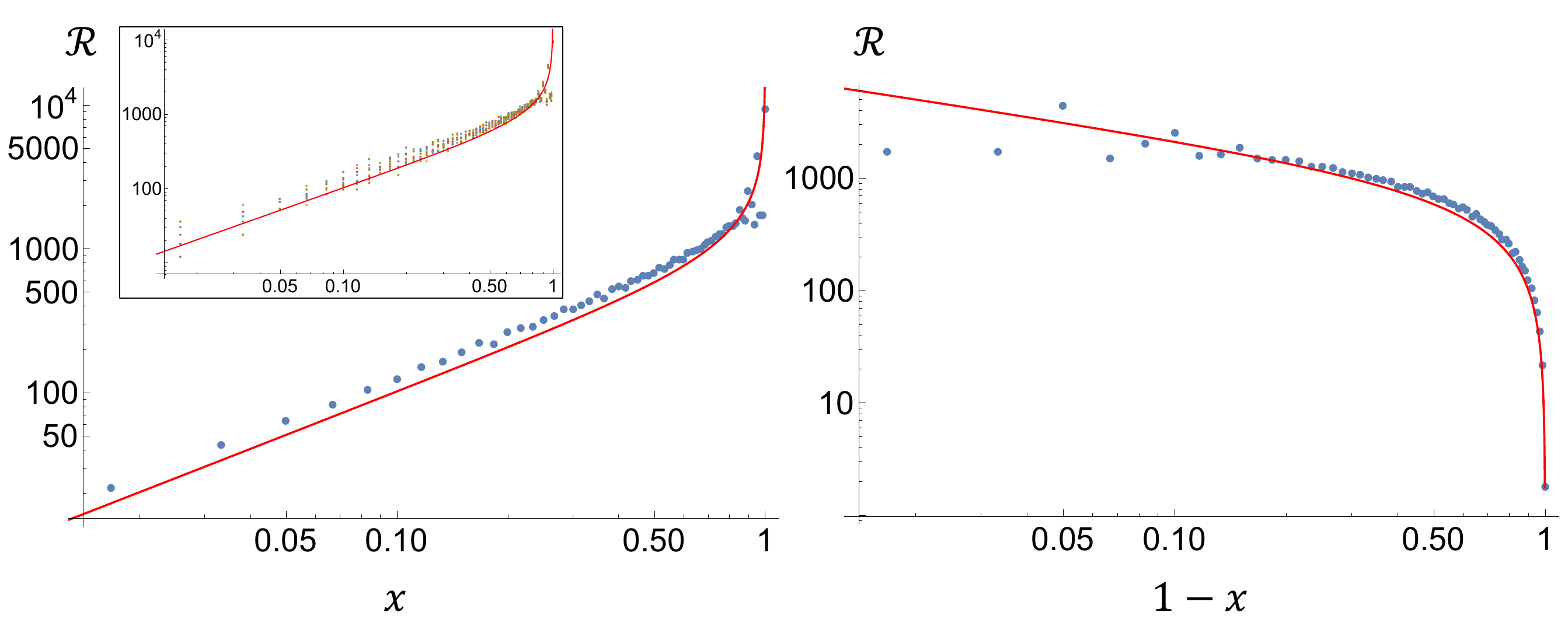}
\caption{Left panel: same as Fig.~\ref{fig:Densities_comparison1}, but on a log-log scale.  Right panel: Same data as the main part of Fig.~\ref{fig:Densities_comparison1}, but plotted versus $1-x$, and on a log-log scale.}
\label{fig:Densities_comparison2}
\end{figure}

We now compare the density profile $\mathcal{R}$ with the data from simulations.  We performed ten simulations from $\tau=17.0$ to $\tau = 17.9$ in increments of $0.1$.  For each value for $\tau$, ten simulations were performed with $N=1000$ particles.  The number of particles within each radius bin were counted (i.e. this will be approximated by $2\pi r \rho(r) \Delta r$, where $\rho(r)$ is the correct continuum radial density), and plotted versus the variable $x = r/R$.  Results over ten simulations for each $\tau$ were averaged.  Inset of Fig.~\ref{fig:Densities_comparison1} shows this average, one for each value of $\tau$.  As the theory suggests, the data (dots) collapses unto one universal curve.  Moreover, this curve closely matches the theoretical $\mathcal{R} = \frac{10}{4} N x \tilde{\rho}(x)$, shown as a red solid curve.  The main part of Fig.~\ref{fig:Densities_comparison1} compares the same theoretical $\mathcal{R}(x)$ with the average of those 10 curves (one for each $\tau$).
The behavior at each end of the curve - one for small $x$ and one for $x$ close to $1$ is elucidated in Fig.~\ref{fig:Densities_comparison2}.  
One feature of Fig.~\ref{fig:Densities_comparison1} deserves a comment.  We note that the dots display oscillatory behavior close to $x=1$.  This happens because in a finite $N$ system, swarmalators organize themselves into rings - see for instance Fig.~\ref{fig:static} or Fig.~\ref{fig:latter_vectors}.  Whether this ring structure remains as $N$ grows is not clear.  Our theory, of course assumes that at sufficiently large $N$ continuum theory will work.  In other words, there exists a continuum density $\rho(\mathbf{x})$, such that the number of swarmalators within a certain area $\Delta A$ that is much smaller than the area of the cluster is accurately given by $\rho(\mathbf{x}) \Delta A$.  This hypothesis is supported by our numerical observations (see for example Fig.~\ref{fig:static}) that the distance between swarmalators becomes smaller relative to the radius as $N$ increases.  The two panels in Fig.~\ref{fig:Densities_comparison2} demonstrate that theory matches numerical simulations quite well. We will now use the continuum theory to make predictions concerning the equilibrium radius and properties of breathing oscillations.

The density profile $\tilde{\rho}$ was obtained by going to the $t \rightarrow \infty$ limit, when all motions ceases.  The system approaches this limit through decaying oscillations: $r_i(t) = r_i^* + \delta r_i(t)$, $R(t) = R^* + \delta R(t)$, and $\theta_i(t) = \Omega t + \delta \theta_i(t)$, where $\delta r_i$, $\delta R(t)$, and $\delta \theta_i(t)$ are decaying functions.  We now introduce the key hypothesis: the profile $\tilde{\rho}$ holds not only at infinite time when oscillations have completely ceased, but even during the mature stages of these decaying oscillations.  What this means is that at rearrangements of particles cease at these latter stages of the oscillations.  We confirmed from simulations that this is indeed the case.  Therefore, the density oscillates because the radius oscillates - the cluster overall expands and contracts, and particles get closer and further apart like on an expanding and contracting rubber sheet, while their relative positions and angles remains the same, and so the functional form of the density profile remains unchanging, i.e. $\tilde{\rho}$ holds true even during these latter stages of the decaying oscillations.  With this key idea in mind, we can now derive coupled equations for the dynamics of $\theta_i$, $r_i$ and $R$.  

As before, we will assume that all $\theta(t)$ are the same (modulo $2\pi$).  This is supported by simulation results - see Fig.~\ref{fig:PhasesOne} and Fig.~\ref{fig:PhasesTwo} for example.  Using the same procedure for passing into the continuum description as before (see discussion preceding Fig.~\ref{fig:six}).  We have \begin{eqnarray}
\dot{\theta} &=& \frac{K}{N} \sin{(\theta(t-\tau) - \theta(t))}\sum_{j \neq i} \frac{1}{\left| \vec{r}_j - \vec{r}_i\right|} \nonumber \\
&=& \frac{2K}{N} \sin{(\theta(t-\tau) - \theta(t))} \int_0^R r \rho(r) dr \int_0^{\pi} \frac{1}{\sqrt{r^2 + l^2 -2rl \cos{\alpha}}}d\alpha \nonumber\\
&=& \frac{2K}{N} \sin{(\theta(t-\tau) - \theta(t))} \int_0^R r \rho(r) \frac{2K\left[\frac{4lr}{(r+l)^2}\right]}{r+l} \,dr 
\end{eqnarray}
We remind the reader that $l$ is radius position of $i$th swarmalator, i.e. it is a constant in our integrals, while $r$ is the radius of $j$th swarmalators over which the summation (or integration, in continuum approximation) is performed.  We now substitute $\rho = \frac{5}{4}\frac{N}{\pi R^2(t)} \tilde{\rho}(r)$. We get
\begin{eqnarray}
\dot{\theta} &=& \frac{10K}{4 \pi R^2(t)} \sin{(\theta(t-\tau) - \theta(t))} \int_0^{R(t)} r\tilde{\rho}(r) \frac{2K\left[\frac{4lr}{(r+l)^2}\right]}{r+l} \,dr \nonumber \\
&=&\frac{10K}{2 \pi R(t)} \sin{(\theta(t-\tau) - \theta(t))} \int_0^1 x\tilde{\rho}(x) \frac{K\left[\frac{4Lx}{(x+L)^2}\right]}{x+L} \,dx,
\end{eqnarray}
where, again $L = l/R$ and $x = r/R$.  But this is precisely the integral that defines $\tilde{\rho}$, and it evaluates to $1$ (see Eq.~\ref{eq:rho-tilde_equation}).  Therefore, our equation becomes
\begin{equation}
\label{eq:theta-equation}
\dot{\theta} = \frac{5K}{\pi R(t)}\sin{(\theta(t-\tau) - \theta(t))} 
\end{equation}
This is an interesting equation, but it is not closed, as we also need an equation for the $R(t)$.

\begin{figure}[H]
\center \includegraphics[width=4in]{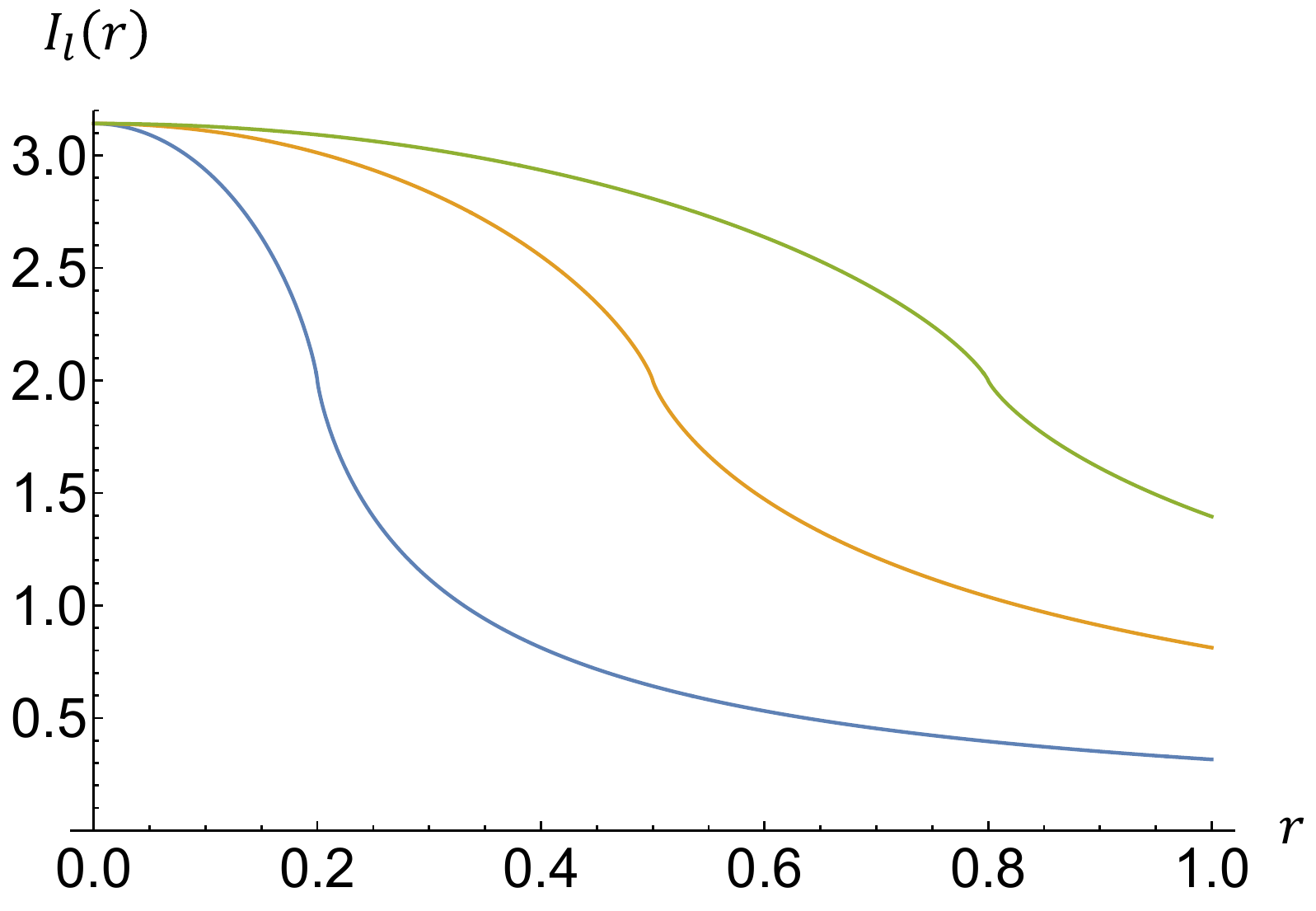}
\caption{$I_l(r)$ for $l=0.2$ (left), $0.5$ (middle) and $0.8$ (right).}
\label{fig:Kernel_function2}
\end{figure}

The spatial equation tells us that
\begin{equation}
\label{eq:radial_dynamics_eq}
\dot{\vec{r}}_i = \frac{1}{N}\sum_{j\neq i}^{N} \left[\frac{\vec{r}_j - \vec{r}_i}{\left|\vec{r}_j - \vec{r}_i\right|}\Big(1+J\cos{\left(\theta(t-\tau) - \theta(t)\right)}\Big) - \frac{\vec{r}_j - \vec{r}_i}{\left|\vec{r}_j - \vec{r}_i\right|^2} \right]
\end{equation}
Consider the first term.  Here, we are adding unit vectors pointing from the $i$-th particle to all other particles.  
Because of circular symmetry, for each particle above, there's another symmetric partner on the opposite side.
\begin{figure}[H]
\center \includegraphics[width=3in]{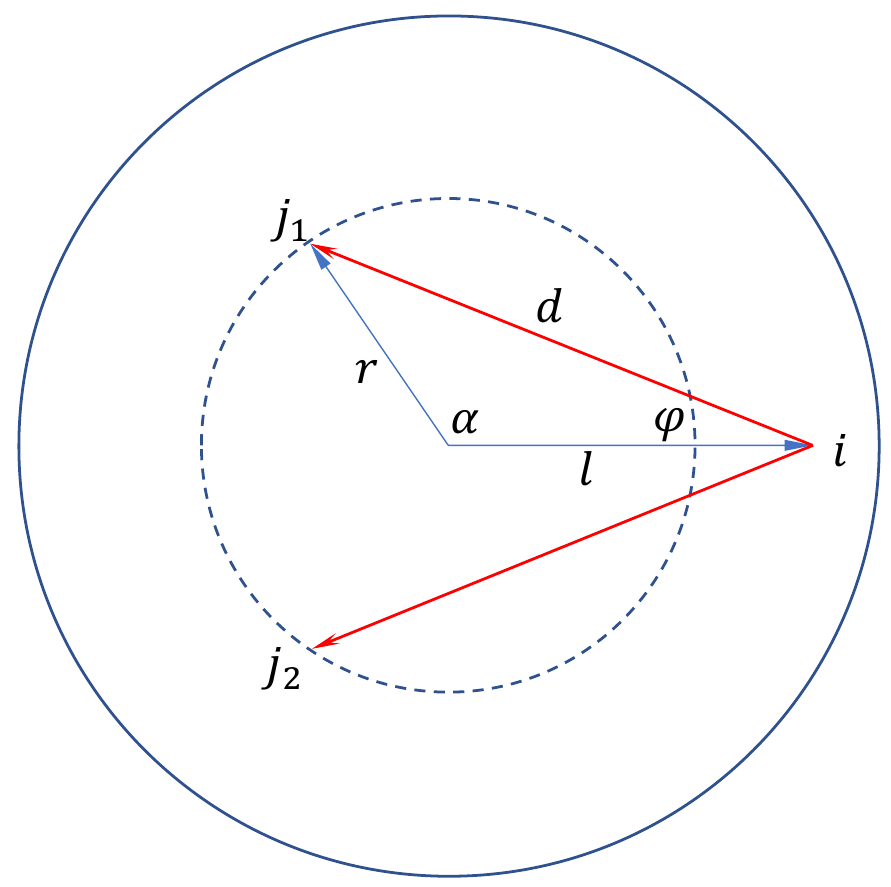}
\label{fig:seven}
\caption{Geometry for evaluating Eq.~(\ref{eq:radial_dynamics_eq}).}
\end{figure}
Thus, for each $i$-th particle, all vectors will point towards the center of the cluster.  Going into the continuum limit, the first sum becomes 
\begin{equation}
-\left[2\Big(1+J\cos{\left(\theta(t-\tau) - \theta(t)\right)}\Big)  \int_0^R r\rho(r) \,dr \int_0^{\pi} \cos{(\varphi(\alpha))} \,d\alpha\right] \hat{R}
\end{equation}
We can find $\cos{(\varphi(\alpha)))}$ using a combination of laws of sines and cosines.  From the law of sines we have
$\sin{\varphi} = \frac{r\sin{\alpha}}{d}$, so $1-\cos^{2}{\varphi} = \frac{r^2\sin^2{\alpha}}{d^2}$.  Thus $\cos{(\varphi(\alpha))} = \sqrt{1-\frac{r^2\sin^2{\alpha}}{r^2 + l^2 -2rl\cos{\alpha}}} = \frac{l-r\cos{\alpha}}{\sqrt{r^2 + l^2 -2rl\cos{\alpha}}}$.   Thus, the inner integral over $\alpha$ is
$\int_0^{\pi} \frac{l-r\cos{\alpha}}{\sqrt{r^2 + l^2 -2rl\cos{\alpha}}}\,d\alpha$.  This integral can be evaluated
\begin{equation}
\label{eq:elliptic-difference}
I_l(r) = \int_0^{\pi} \frac{l-r\cos{\alpha}}{\sqrt{r^2 + l^2 -2rl\cos{\alpha}}}\,d\alpha = \left|\frac{(r-l)}{l}E\left[-\frac{4rl}{(r-l)^2}\right] - \frac{(r+l)}{l}K\left[-\frac{4rl}{(r-l)^2}\right] \right|,
\end{equation}
where $E$ is a complete elliptic integral of the second kind.  We plot $I_l(r)$ for several $l$ in Fig.~\ref{fig:Kernel_function2}.
Note that although $I_l(r)$ quantity is not defined at $r=l$, the limit of $r\rightarrow l$ from both sides exists and equals $2$.  Thus, 
\small
\begin{equation}
\mbox{The first term in Eq.~(\ref{eq:radial_dynamics_eq})} = -\left[2\Big(1+J\cos{\left(\theta(t-\tau) - \theta(t)\right)}\Big)  \int_0^R r\rho(r) \left|\frac{(r-l)}{l}E\left[-\frac{4rl}{(r-l)^2}\right] - \frac{(r+l)}{l}K\left[-\frac{4rl}{(r-l)^2}\right] \right| \,dr\right] \hat{R}
\end{equation}
\normalsize
We now turn attention to the second term.  The inner integral in this term would be $2\int_0^{\pi} \frac{l-r\cos{\alpha}}{r^2 + l^2 -2rl\cos{\alpha}}\,d\alpha$.  It evaluates to $2\pi/l$ when $r<l$ and $0$ when $r>l$.  In the second case, the repulsive force from particles ``to the right'' of the $i$-th particle and the repulsive force from particles ``to the left'' of the $i$-th particle add up to zero.  This is similar to the electric (or gravitational) field inside a hollow shell.  A particle inside a shell of charge (or mass) experiences no net force.  But a particle outside a sphere of charge (or mass) does experience net force.  So, the second term in the spatial equation becomes $\left[\frac{2\pi}{l}\int_0^l r\rho(r) \,dr\right]\hat{R}$.  All together, the spatial equation says
\small
\begin{equation}
\dot{l} = \frac{1}{N}\left[-2\Big(1+J\cos{\left(\theta(t-\tau) - \theta(t)\right)}\Big)\int_0^R r\rho(r)  \left|\frac{(r-l)}{l}E\left[-\frac{4rl}{(r-l)^2}\right] - \frac{(r+l)}{l}K\left[-\frac{4rl}{(r-l)^2}\right] \right| \,dr  + \frac{2\pi}{l}\int_0^l r\rho(r) \,dr\right],
\end{equation}
\normalsize
and no dynamics in the angular direction.  Evidently, the circular symmetry leads to the prediction that all particle motion is going to be in the radial direction. 
This can only be expected to be true in the continuum limit; in the discrete case, there might not be an exact cancellation of the two vectors, which would allow for some angular motion.  

We now substitute for $\rho(r) = \frac{5}{4}\frac{N}{\pi R^2} \tilde{\rho}(r)$ (see Eq.~(\ref{eq:Rho_Model})), using $\tilde{\rho} = 0.3(1-r/R)^{-1/2}$, which, as we saw, is a good model of the numerical solution.  The first integral can't be expressed in terms of elementary functions, but it is very closely approximated by $\frac{5}{4}\frac{N}{\pi} \frac{l}{R}$.  The second integral evaluates to $\frac{5}{4}\frac{N}{\pi} \times 0.3\left(\frac{4}{3} -\frac{2(2+l/R)}{3}\sqrt{1-l/R}\right)$.  Therefore,
\begin{equation}
\label{eq:key_result}
\dot{l} = \frac{5}{2 \pi}\left[-\Big(1+J\cos{\left(\theta(t-\tau) - \theta(t)\right)}\Big)\left(\frac{l}{R}\right)  + \frac{0.3\pi}{l} \left(\frac{4}{3} -\frac{2(2+l/R)}{3}\sqrt{1-l/R}\right)\right].
\end{equation}
Three assumptions that went into this result are: (i) circular symmetry, (ii) that $\tilde{\rho}$ works even during the latter stages of decaying oscillations, when particle rearrangements have ceased, and (iii) that all $\dot{\theta}_i$ are identical.  All three assumptions are corroborated by simulations. 

\begin{figure}[H]
\center \includegraphics[width=7in]{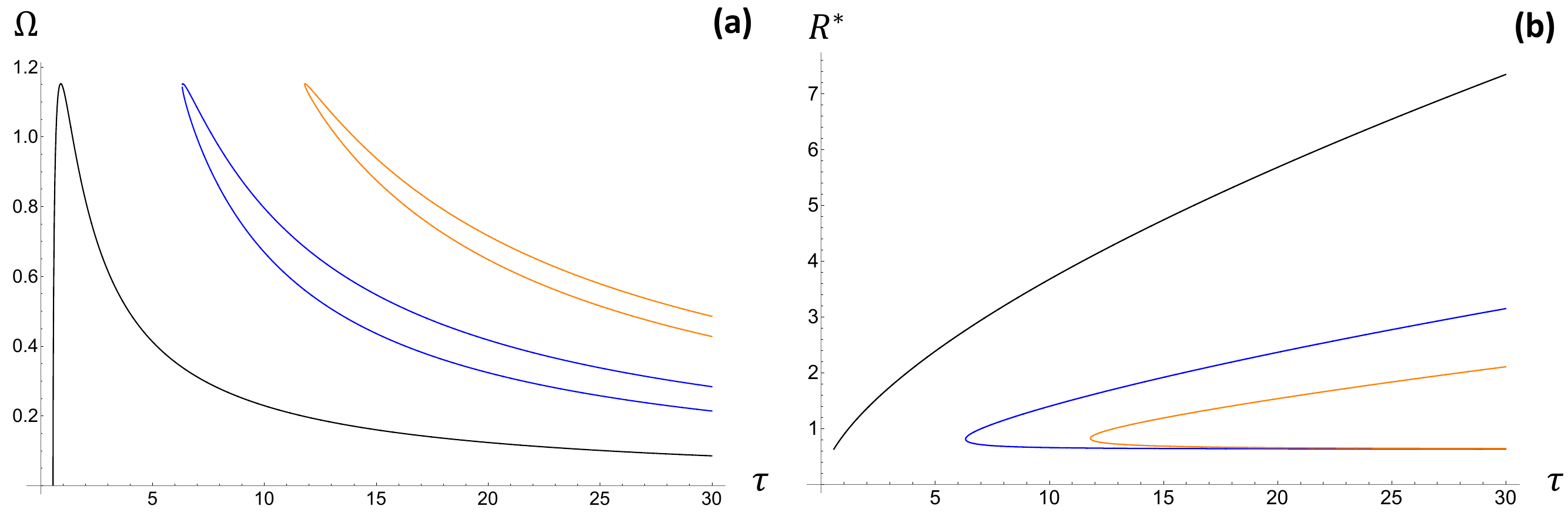}
\caption{(a) $\Omega(\tau)$.  (b) $R^*(\tau)$.  The parameters are $J=1$, $K=-0.7$.}
\label{fig:theory_equil}
\end{figure}

Eq.~(\ref{eq:key_result}) gives the instantaneous radial velocity of a particle located a distance $l$ from the center.  Setting $l=R$, gives the velocity of a particle on the edge, i.e. it gives $\dot{R}$.  Thus, we finally arrive at two coupled equations for dynamics of $R$ and $\theta$ (we reproduce here Eq.~(\ref{eq:theta-equation}) for completeness):
\begin{eqnarray}
\label{eq:theta-equation_again}
\dot{\theta} &=& \frac{5K}{\pi R(t)}\sin{(\theta(t-\tau) - \theta(t))} \\
\label{eq:r-equation}
\dot{R} &=& \frac{5}{2\pi}\left[-\Big(1+J\cos{\left(\theta(t-\tau) - \theta(t)\right)}\Big) + \frac{2}{5}\frac{\pi}{R}\right]
\end{eqnarray}
Equations (\ref{eq:theta-equation_again}) and (\ref{eq:r-equation}) are two coupled equations for $\theta(t)$ and $R(t)$.  We will study the dynamics soon, but first, we analyze the $t\rightarrow \infty$ state, i.e. $R=R^*$ - the constant value into which the radius settles, and $\theta = \Omega t$.  These will be functions of $J$, $K$, and $\tau$ as we wanted.  The following must be true in this static limit:
\begin{eqnarray}
\label{eq:eq1} R^* &=& -\frac{5K}{\pi}\frac{\sin{(\Omega \tau)}}{\Omega}  \\
\label{eq:eq2} R^* &=& \frac{2\pi/5}{1+J\cos{(\Omega \tau)}}
\end{eqnarray}
We have already encountered the first of these - see Eq.~(\ref{eq:R-Omega_1}).  Combining the two, we get
\begin{equation}
\label{eq:Omega-eqn}
1+J\cos{(\Omega \tau)} = -\frac{2\pi^2}{25} \frac{\Omega/K}{\sin{(\Omega \tau)}}
\end{equation}
The solution $\Omega(\tau)$ is plotted in Fig.~\ref{fig:theory_equil}(a).  Evidently, the solutions are multi-valued.  The resulting $R^*(\tau)$, obtained through Eq.~(\ref{eq:eq1}) is shown in Fig.~\ref{fig:theory_equil}(b).  
%
%
We learn that theory predicts multiple stable states.  Focusing on the first one (left-most, black curves in Fig.~\ref{fig:theory_equil}) there is a transition point $\tau_l$ below which $\Omega=0$.  Note that below this $\tau_l$, the steady state condition of Eq.~(\ref{eq:theta-equation_again}) simply says $0=0$, and equilibrium radius is determined only from Eq.~(\ref{eq:r-equation}) (or from Eq.~(\ref{eq:eq2})), giving $R^*_l=\frac{2\pi/5}{1+J}$.  The expression for $\tau_l$ can be obtained, for instance, by combining this result with Eq.~(\ref{eq:Omega-eqn}) with $\Omega=0$, giving $\tau_l = -\frac{2\pi^2}{25} \frac{1}{K(1+J)}$.

Above $\tau_l$, our continuum theory predicts a well-defined value of $R^*$.  This is different from simulations, where $R^*$ is only truly well-defined above $\tau_c > \tau_l$, although one can meaningfully talk about a time averaged value of the radius of the cluster even below $\tau_c$ (recall that $\tau_c$ denotes a transition between two types of long-time behaviors - from boiling to quasi-static pseudo-crystal, while $\tau_l$ denotes a transition in the dynamics - from synchronized with a non-zero average $\dot{\theta}$ to unsynchronized with a zero average $\dot{\theta}$).  On the other hand, the concept of $\tau_c$ does not exist in the continuum theory, because this theory is oblivious to surface fluctuations. 

\begin{figure}[H]
\center \includegraphics[width=5in]{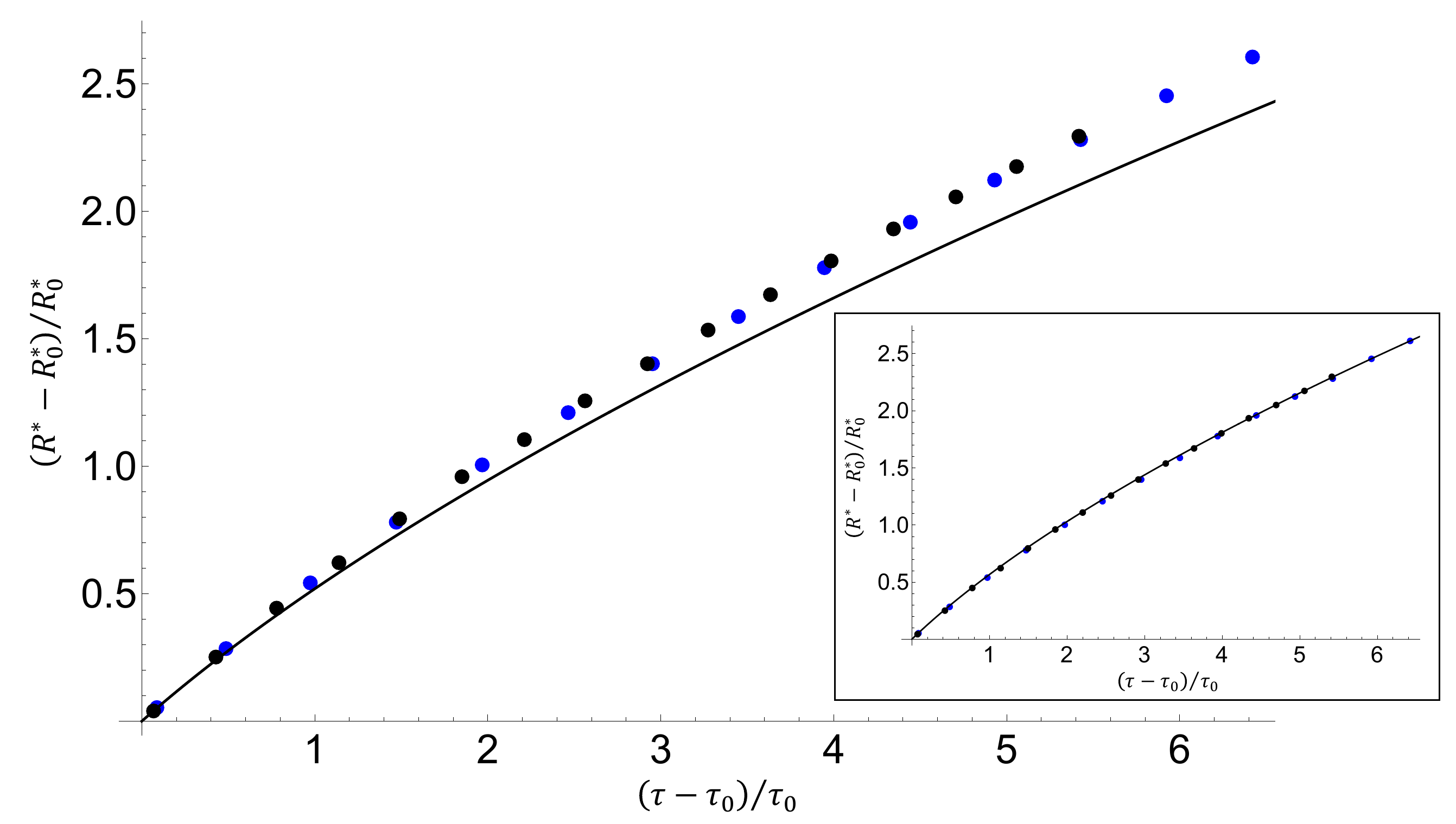}
\caption{The dots represent the simulation data (same data as in Fig.~\ref{fig:R*_vs_tau}(b)).  Solid line is the theoretical prediction.  Here $\tau_0$ is $\tau_c$ for simulation results and $\tau_l = -\frac{2\pi^2}{25}\frac{1}{K(1+J)}$ for continuum theory.  Similarly, $R_0^*$ is $R_c^*$ for simulation results and $R_l^* = \frac{2\pi/5}{1 + J}$ for continuum theory.  The difference between the two predictions disappears if the theoretical $y-$values are multiplied by a factor $\approx 1.09$, as shown in the inset. The parameters are $(J=1,K=-0.7)$.}
\label{fig:theory_vs_sim_equil}
\end{figure}

The shape of $R^*(\tau)$ in the first branch of Fig.~\ref{fig:theory_equil} strongly resembles the shape of $R^*(\tau)$ obtained from simulations (see Fig.~\ref{fig:R*_vs_tau}).  We now would like  to compare the two predictions.  Note that the theoretical prediction is independent of $N$, while in presenting simulation results we observed the data collapse over different system sizes when instead of plotting $R^*$ vs.~$\tau$ we plotted the dimensionless $(R^* - R^*_c)/R^*_c$ vs.~$(\tau-\tau_c)/\tau_c$.  Equilibrium radius only truly makes sense above $\tau_c$.  This justifies why the rescaling had to be done with respect to $\tau_c$, rather than $\tau_l$. On the other hand, $\tau_c$ does not exist in the continuum theory, while the cluster radius is a function that is a constant ($\frac{2\pi/5}{1+J}$) below $\tau_l$, and grows above $\tau_l$.  For this reason, we rescaled the $x$-axis of the theoretical prediction by $\tau_l$ and the $y$-axis by $(2\pi/5)/(1 + J)$, and this was then compared with the dimensionless simulation data.  This is shown in Fig.~\ref{fig:theory_vs_sim_equil}.
While there is about a $9\%$ difference between the predictions of the continuum theory with simulations, they in fact appear to match in functional form.  This is seen in the inset of Fig.~\ref{fig:theory_vs_sim_equil}, where we multiplied the value of the analytical function by $1.09$.  

\begin{figure}[H]
\center \includegraphics[width=7in]{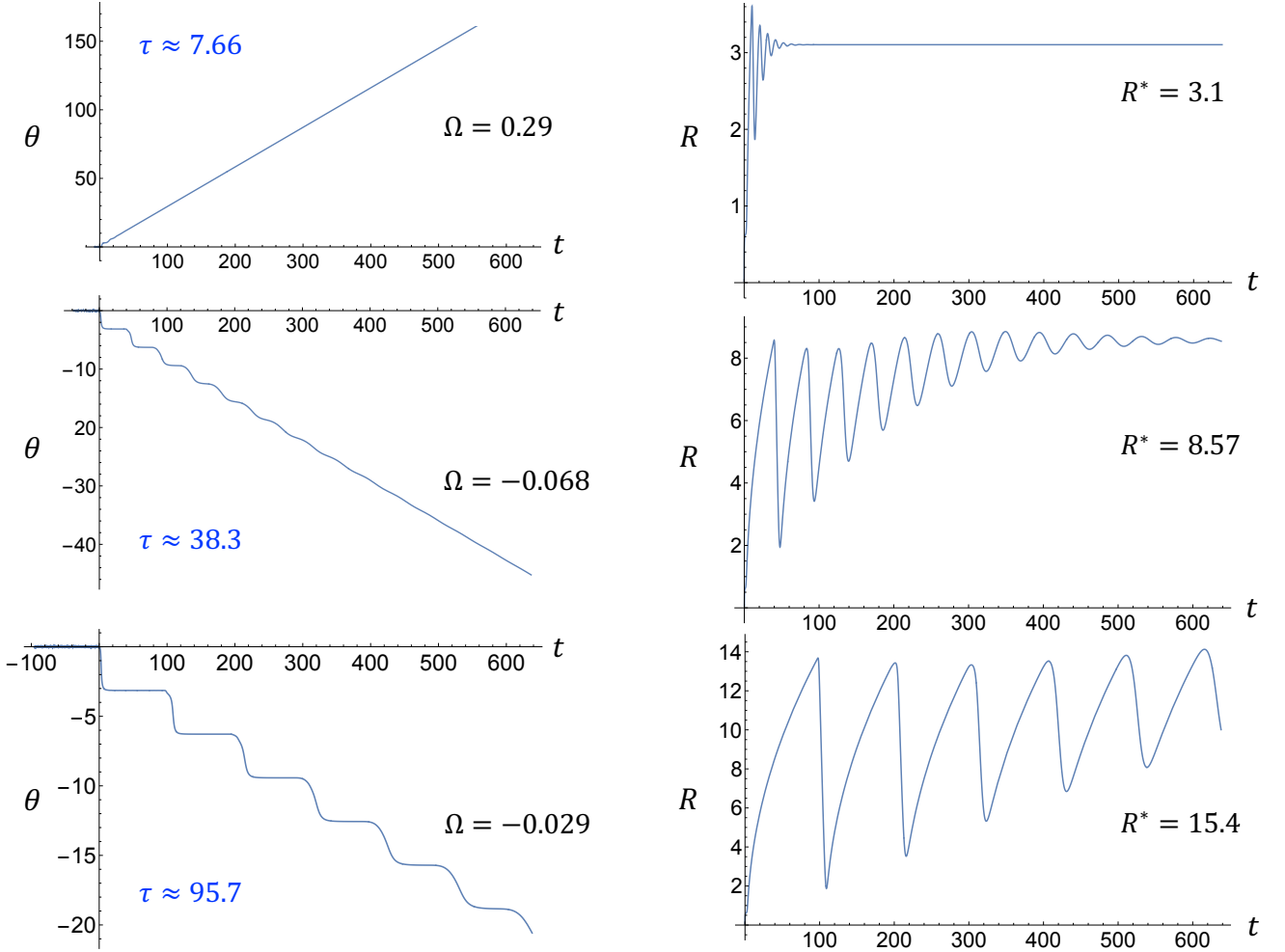}
\caption{Evolution of $\theta(t)$ and $R(t)$ with increasing $\tau$, as predicted by the continuum theory.  Here $(J=1, K=-0.7)$.  This gives $\tau_l \approx 0.564$.  The values of $\tau$ were chosen to be the same fraction of $\tau_l$ as in Fig.~\ref{fig:dyn_sim_exs} that depicts simulation results (this assumes that $\tau_l$ in Fig.~\ref{fig:dyn_sim_exs} is $1.65$ - see caption of that figure).  Also, the evaluation time ($638$) in these figures was chosen to be the same fraction of $\tau_l$ as in Fig.~\ref{fig:dyn_sim_exs} , namely $\approx 1131.3 \tau_l$.}
\label{fig:dyn_theory_exs}
\end{figure}

We now turn attention to relaxational dynamics of breathing oscillations predicted by Eqs.~(\ref{eq:theta-equation_again}) and (\ref{eq:r-equation}).  Fig.~\ref{fig:dyn_theory_exs} depicts some examples of the evolution of $\theta$ and $R$ in time.  Solving these delayed equations numerically - as in solving the full dynamical equations - requires prehistory conditions.  However, in contrast to the simulations of dynamics from the full set of equations, here a constant initial condition would predict no evolution: $\theta(-\tau) - \theta(0) = 0$ results in $\dot{\theta}=0$ and $\dot{R}=0$ at $t=0$.  Our theory is meant to produce long time asymptotic behavior.  Therefore, we chose prehistory that simulates complex, random-like behavior at earlier times (we see from simulations that early transient is complex).  To this extent, we used the following prehistory for $-\tau<t<0$:
\begin{eqnarray}
\theta(t) &=& 0.01\sum_{n=1}^{100} a_n \sin{(nt - n^2)} \\
R(t) &=& 0.01\sum_{n=1}^{100} a_n \cos{(nt - n^2)},
\end{eqnarray}
where ${a_n}$ were a set of random numbers between $0$ and $1$.  The phase offsets were used to eliminate spikes when $t$ is an integer multiple of $2\pi$.  This is the type of prehistory that was used in producing Fig.~\ref{fig:dyn_theory_exs}.  
The step-wise shape of $\theta(t)$, as well as the shape of $R(t)$ is qualitatively the same as simulations (see Fig.~\ref{fig:dyn_sim_exs} below). We see that as $\tau$ increases, the period of decaying oscillations increases, and the decay rate decreases.  These qualitative observations also match simulations. As time grows, both $\delta \theta(t)$ and $\delta R(t)$ evolve to a more pure harmonic, i.e. higher harmonics decay away quicker.

In presenting simulation results, we focused on the dynamical properties of long-time relaxations towards the equilibrium.  There are only two:  frequency and decay rate.  To extract these from Eqs.~(\ref{eq:theta-equation_again}) and (\ref{eq:r-equation}) we linearize them by setting $\theta(t) = \Omega t + \delta \theta$ and $R(t) = R^* + \delta R$.  The resulting linear equations for $\delta \theta$ and $\delta R$ are
\begin{eqnarray}
\label{eq:lin_theta} \frac{d}{dt} \delta \theta &=&  \left(\delta \theta(t-\tau) - \delta \theta(t)\right)\left(\frac{5K\cos{(\Omega \tau)}}{\pi R^*}\right) + \delta R(t) \left(\frac{5K}{\pi \left(R^*\right)^2}\sin{(\Omega \tau)}\right)\\
\label{eq:lin_R}\frac{d}{dt} \delta R &=& -\frac{5}{2 \pi}J\sin{(\Omega \tau)}\left(\delta \theta(t-\tau) - \delta \theta(t)\right) - \frac{1}{\left(R^*\right)^2} \delta R
\end{eqnarray}
Next, we seek solutions in the form 
\begin{equation}
\left(\begin{array}{c} \delta \theta \\ \delta R \end{array}\right) = \left(\begin{array}{c} \delta \theta_0 \\ \delta R_0 \end{array}\right) e^{-\lambda t}
\end{equation}
Substituting this ansatz into Eqs.~(\ref{eq:lin_theta})-(\ref{eq:lin_R}) gives the following equation for eigenvalue $\lambda$:
\begin{equation}
\det \left(\begin{array}{cc} \left(e^{\lambda \tau} - 1\right)\left(\frac{5K\cos{(\Omega \tau)}}{\pi R^*}\right) + \lambda & \frac{5K \sin{(\Omega \tau)}}{\pi \left(R^*\right)^2} \\
-\left(e^{\lambda \tau} - 1\right)\left(\frac{5J}{2 \pi} \sin{(\Omega \tau)}\right) & -\frac{1}{\left(R^*\right)^2} + \lambda\end{array}\right) = 0
\end{equation}
The solutions are generally complex.  When we set $\lambda = r + i\omega$, substitute into the above equation, compute the determinant, and separate the real and imaginary parts, we get a pair of equations for two variables $r$ and $\omega$.  Each equation can be represented graphically as a zero contour of a function of two variables $r$ and $\omega$.  The solutions take place at the intersection of the two sets of contours.  There is an infinite number of solutions, and we performed a numerical search for the solution with the lowest real part, which dominates at large times.  For $\tau$ sufficiently close to $\tau_l$ the $\lambda$ with the lowest real part is real and the motion becomes overdamped.  Above such $\tau$, the $\lambda$ gains an imaginary part, and we get a complex conjugate pair of solutions.

\begin{figure}[H]
\center \includegraphics[width=5.5in]{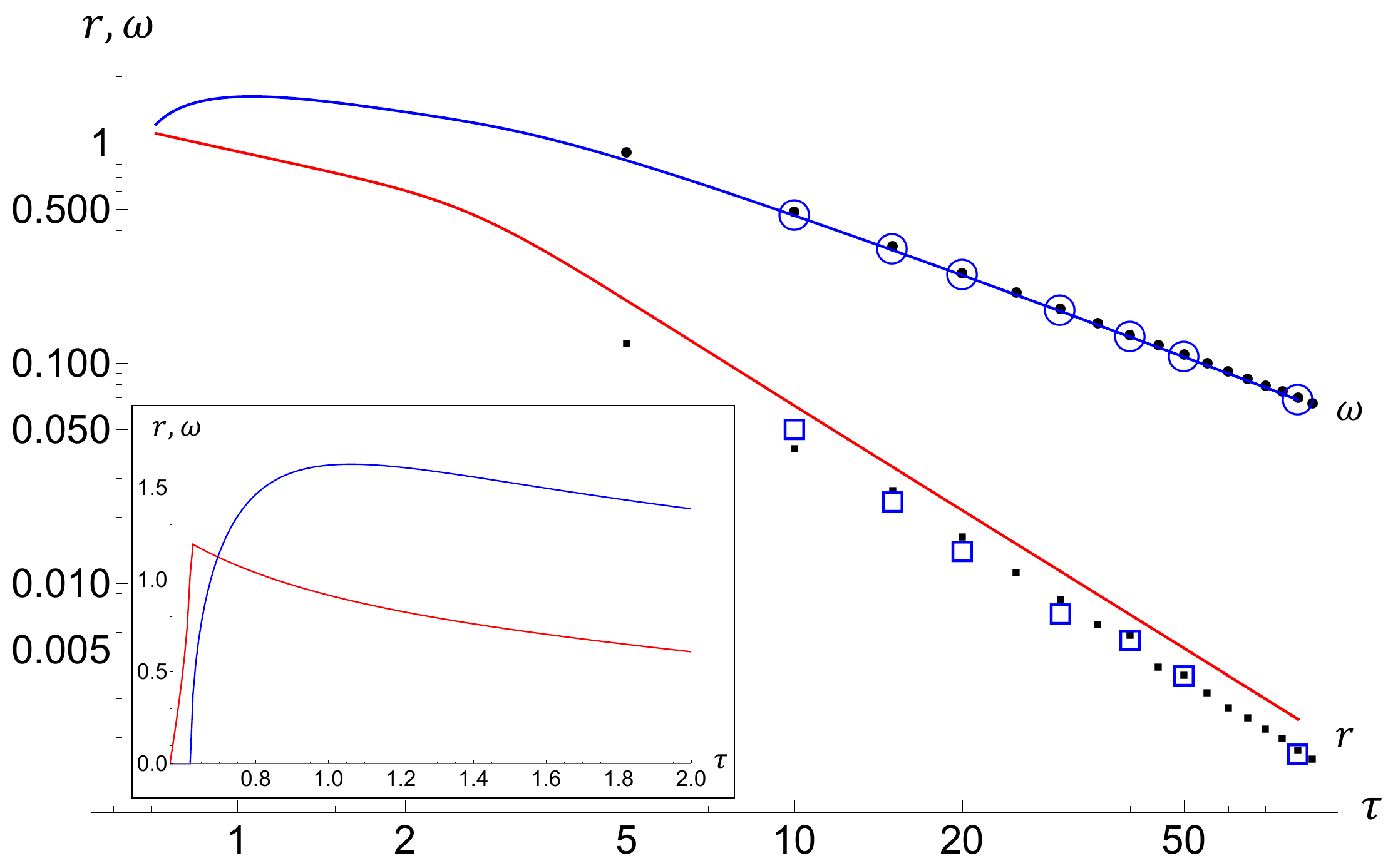}
\caption{The numerical data (squares and circles) is exactly the same as in Fig.~\ref{fig:rates_omegas_numerics}.  The solid lines are theoretical predictions.  Decay rate $r$ (lower, red) and angular frequency $\omega$ (upper, blue).  Inset: theoretical $r$ and $\omega$ at low $\tau>\tau_l$, plotted on a linear scale.  The transition to overdamped solutions, where eigenvalues are purely real (i.e. $\omega=0$) is clearly seen at $\tau \approx 0.62$.  The values of $\tau_c$ are $10.11 \pm 0.24$ for $N=300$  (blue open symbols) and $14.03 \pm 0.4$ (black solid symbols).}
\label{fig:rates_omegas_compare}
\end{figure}

In presenting dynamical properties of oscillations (for example, $\omega$ and $r$) we observed that unlike the static properties, it was not necessary to perform rescaling of variables - the data for various $N$ already collapses.  This suggest that in comparing theory with simulations we will also not perform rescaling for the theoretical functions, such as $r(\tau)$ or $\omega(\tau)$.  With this idea in mind, we now compare these theoretical predictions with simulation results.  This comparison is shown in Fig.~\ref{fig:rates_omegas_compare}.  The theory (solid lines) predicts the scaling of both  $\omega(\tau)$ and $r(\tau)$ very well.  In the case of $\omega$s, it predicts the actual values, while there is a small overall multiplicative factor for decay rates. 

In comparison between the predictions of the reduced-order theory and full simulations we see a very good match in the asymptotic scaling of $r(\tau)$ and $\Omega(\tau)$ at large $\tau$.  We remind the reader that it was not possible to obtain simulation results for sufficiently low $\tau$, and the reason for this is explained at the end of Section \ref{sec:transitions}.  However, we can compare the values of $\tau_l$, where phase synchronization first appears.  The value of $\tau_l$ from the continuum theory is $\approx 0.564$ for $(J=1, K=-0.7)$.   The value of $\tau_l$ from simulations for the same $J$ and $K$ was found to be between $1.6$ and $1.7$ for $N=300$ and $N=400$, and between $1.7$ and $1.8$ for $N=100$.  As $\tau$ is lowered significantly below $\tau_c$, the thickness of the boiling layer grows (see Fig.~\ref{fig:moustail_evolution}).  In this regime there is a strong variation of velocity vectors between even nearest neighbor particles.  Therefore, we expect the continuum theory to break down for $\tau$ significantly below $\tau_c$.  Notice that there is still a good match below, but close to $\tau_c$.

\section{Discussion and summary}
\label{sec:discission}
We presented the first study on the role of time delay in interacting swarmalators, although a recent work has studied delayed in vicsek type models \cite{sun2022delay}.  Two new long-time collective states due to delay were discovered.  The first is the quasi-static pseudo-crystal, and the second is the boiling state.  In the first state, swarmalators settle into quasi-static cluster in a pseudo-crystaline arrangement.  Swarmalators execute creeping motions with very small amplitudes.  In the second state that happens at lower values of time delay, the swarmalators close to the surface perform boiling-like convective motions.  The transient that leads into both of these states has an early time component, and a much longer stage that involves collective oscillations of the whole cluster, which give the cluster the breathing-like effect.  Throughout most of this longer phase of the transient particles have already finished rearrangements, and internal phases of swarmalators have already synchronized.  We have not thoroughly mapped the $(J,K)$ parameter space, so other new collective phenomena cause by the delay are possible.

We also proposed a phenomenological continuum theory, based on a key ansatz that particle rearrangements complete at fairly early times, so particles have settled into fixed relative positions during the latter stages of breathing oscillations.  Therefore, the infinite time equilibrium density profile - which we were able to calculate using this continuum theory - also holds throughout the latter stages of the breathing.  This allowed us to calculate frequency and decay rates of breathing, which match numerical results well.  This ansatz is confirmed with simulations, but it would need to be put on a firmer theoretical understanding in future work.  The other two assumptions are circular symmetry and phase synchronization at later stages of the breathing.  The existence of early phase synchronization especially also needs to be understood deeper in the future.  However, all three assumptions are corroborated with numerical simulations.   While the phase slips appear to take place simulataneously for all swarmaaltors, there are tiny differences in which particles experience the slips first.  It would also be interesting to understand if there's a relationship between local structural properties and dynamics of phase slips.  

Quasi-static pseudo-crystal -  such as the creeping motion in the pseudo-static quasi-crystal - reminds us of glassy phenomena.  Exploring the role of frustration and aging on swarmalator phenomena  would be a tantalizing avenue of future research.  Another direction - which we are currently pursuing - is the study of delay in simpler models, such as swarmalators on a ring.

\section{Acknowledgements}
We would like to thank Evgeniy Khain for a useful discussion.  A part of this work was done with the financial support from William and Linda Frost fund for undergraduate summer research at Cal Poly.  An earlier version of this work was a part of an undergraduate senior project by one of the co-authors, Nicholas Blum \cite{blum2021swarming}.  Some of the figures from this senior project are reproduced here.  In submitting the project to Cal Poly Digital Commons, Nicholas Blum entered the agreement that stated the following: ``No copyrights are transferred by this agreement, so I, as the Contributor, or the copyright holder if different from the Contributor, shall retain all rights. If I am the copyright holder of a Submission, I represent that third-party content included in the Submission has been used with permission from the copyright holder(s) or falls within fair use under United States law.''
\newpage
\begin{appendix}
\section{Additional plots}
\label{sec:Details}
\begin{figure}[H]
\center \includegraphics[width=5.8in]{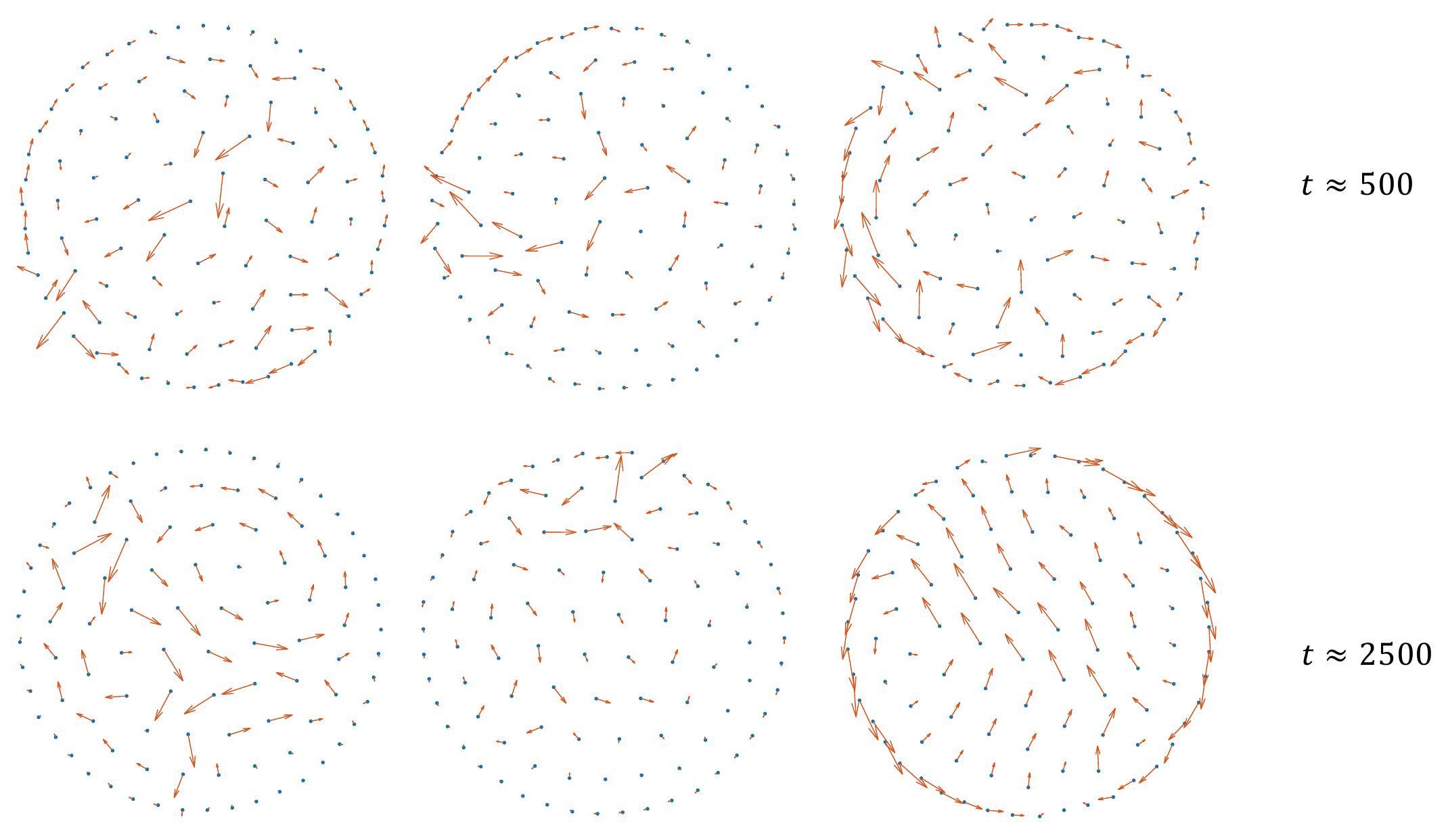}
\caption{$N=100$.  In all examples, breathing ended between $t=250$ and $t=300$.  Here $\tau=1.5 \tau_c$ with $J=1$ and $K=-0.7$.}
\label{fig:N100_mousetail}
\end{figure}

\begin{figure}[H]
\center \includegraphics[width=5.8in]{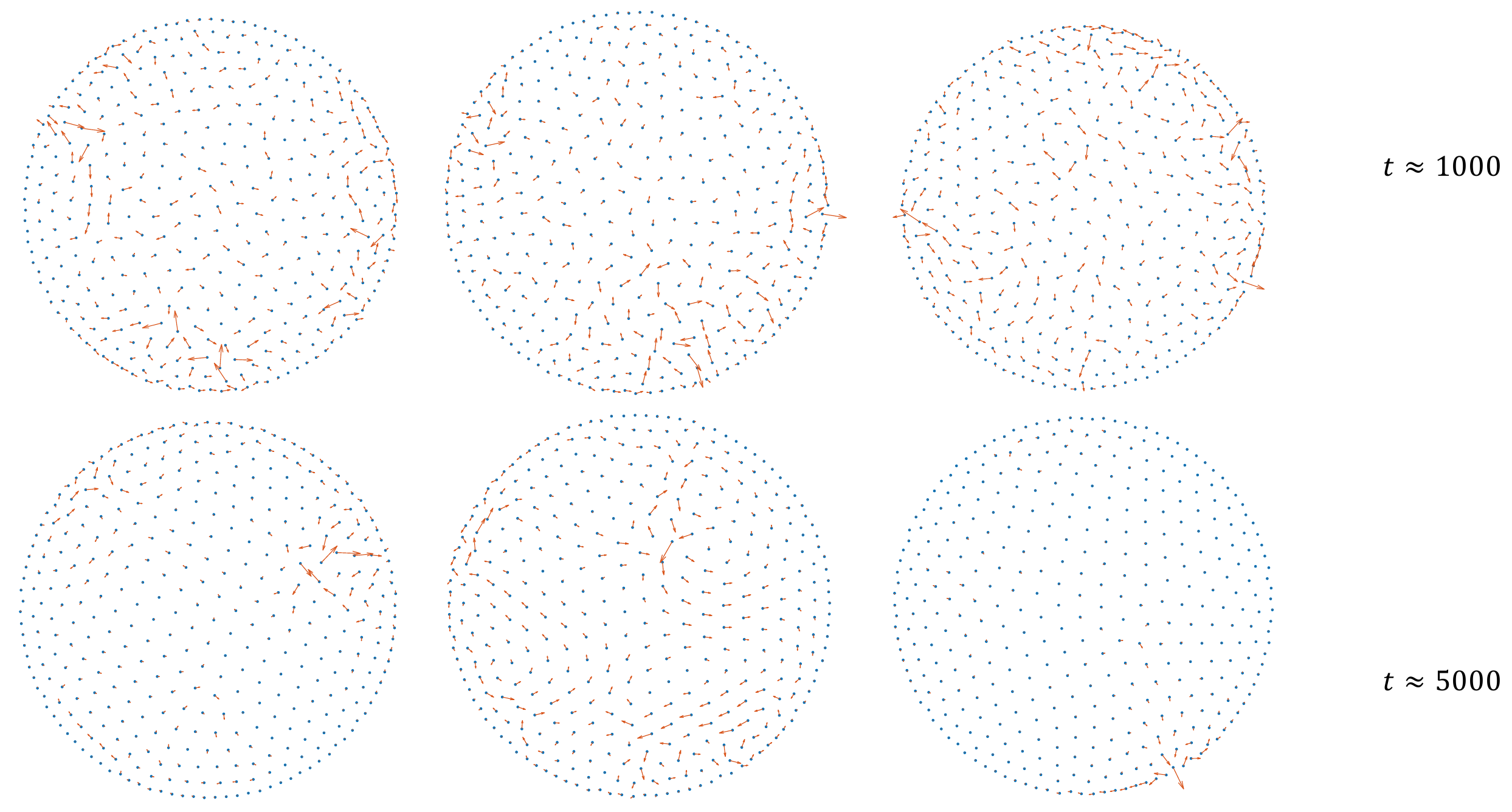}
\caption{$N=400$.  In all examples, breathing ended between $t=500$ and $t=600$.  Here $\tau=1.5 \tau_c$ with $J=1$ and $K=-0.7$.}
\label{fig:N400_mousetail}
\end{figure}

\begin{figure}[H]
\center \includegraphics[width=5.5in]{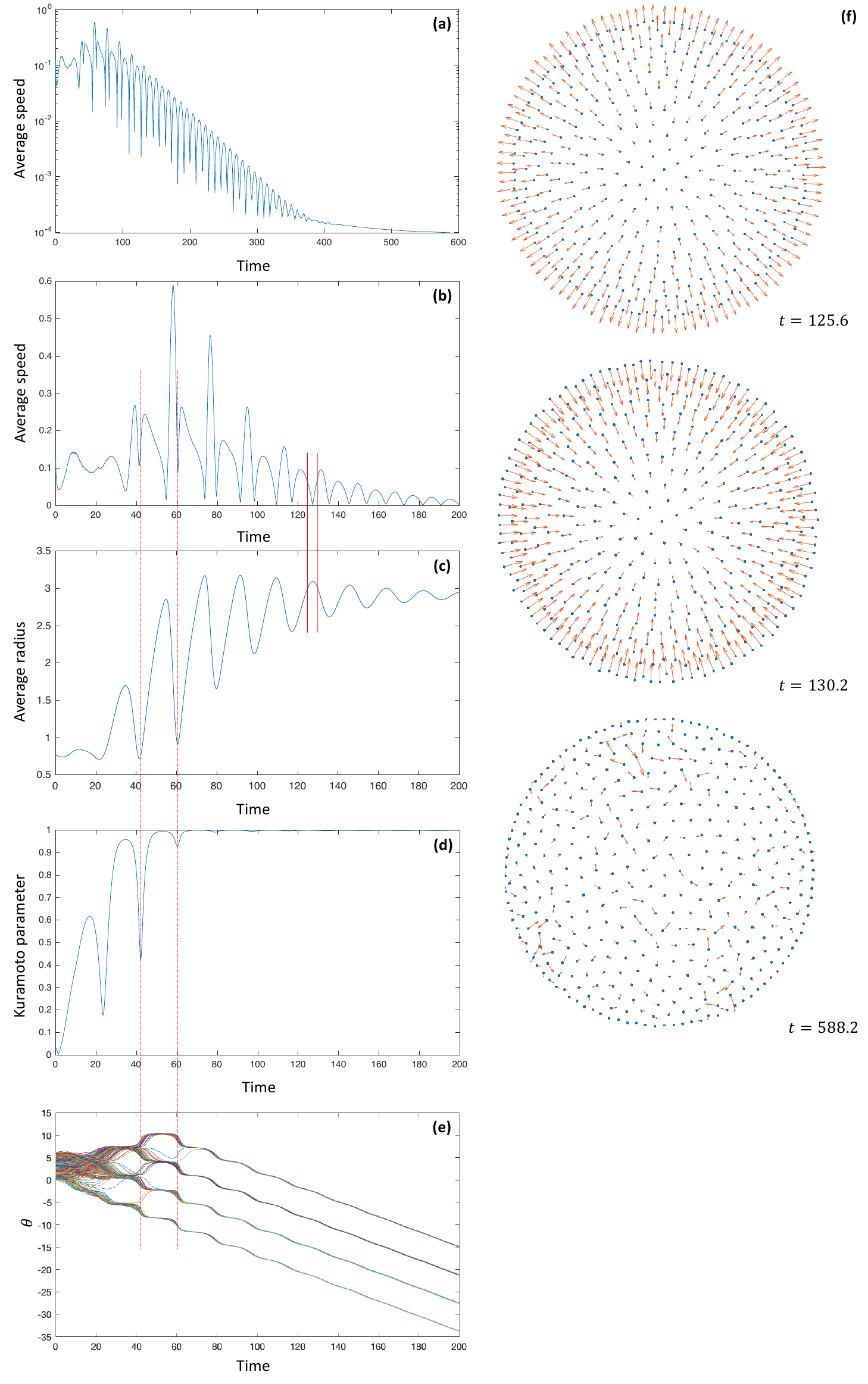}
\caption{$J=1$, $K=-0.7$, $N=400$, $\tau =1.3\tau_c$.  (a) $|\overline{v}|(t)$ on a logarithmic scale.  (b) $|\overline{v}|(t)$ on a linear scale and in a shorter time window that shows only the early transient, the onset of breathing, and breathing in their latter stages.  (c) $\overline{R}(t)$.  (d) Kuramoto order parameter.  (e) $\theta(t)$ for each swarmalator.  (f) Snapshots of particle positions and velocity vectors at three instants of time.}
\label{fig:big_example}
\end{figure}

The following two figures provide more examples of the creeping motion that takes place after the breathing transient.   Fig.~\ref{fig:N100_mousetail} is for $N=100$ and Fig.~\ref{fig:N400_mousetail} is for $N=400$.  The lengths of arrows that represent particle velocity vectors have been up-scaled to be visible; of course these velocities are very small in comparison to velocities during the breathing stage.  We see that in some cases, the velocity patterns are localized, but not always.  It may be that localization is more common at later times, and occurs near the edge, but we have not done a systematic study in order to conclude this definitively.
Next, we present an example of the dynamics for $\tau$ above $\tau_c$ in Fig.~\ref{fig:big_example}.  
The two long vertical red lines are placed at two subsequent dips in the Kutamoto order parameter.  These dips happen at the time when the phases slip.  When this happens, there is a window of time when the phases of swarmalators are not all the same, before they all re-synchronize.  The details of this process can be a subject of study in the future.  We also show several patterns: two during the late stages of breathing (we placed two short red lines in $\overline{v}(t)$ and $R(t)$ graphs at approximately these instants of time), and one post-breathing.
\\

The next plot demonstrates how the dynamics as predicted by simulations evolves with $\tau$.
\begin{figure}[H]
\center \includegraphics[width=7.2in]{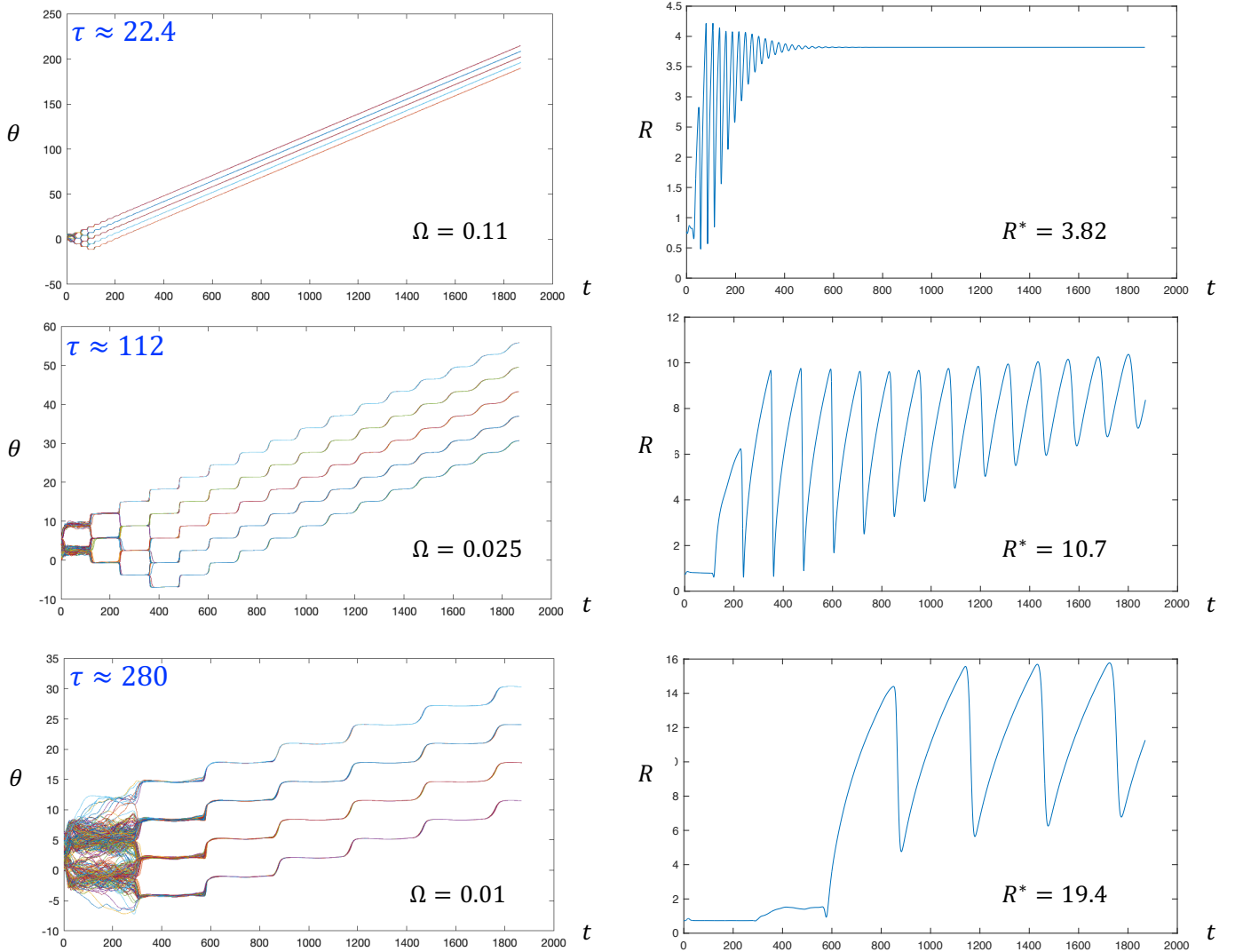}
\caption{Evolution of $\theta(t)$ and $R(t)$ with increasing $\tau$, as predicted by the simulations.  Here $K=-0.7$ and $N=400$.  For this $(N,K)$, $\tau_c =11.2 \pm 0.26$, so the first graph is at $\tau = 2\tau_c$.  The value of $\tau_l$ is between $1.6$ and $1.7$.}
\label{fig:dyn_sim_exs}
\end{figure}
The next plot demonstrates how the long-time behavior that takes places after breathing oscillations evolves with $\tau$, as it is lowered from above to below $\tau_c$.  
\begin{figure}[H]
\center \includegraphics[width=5in]{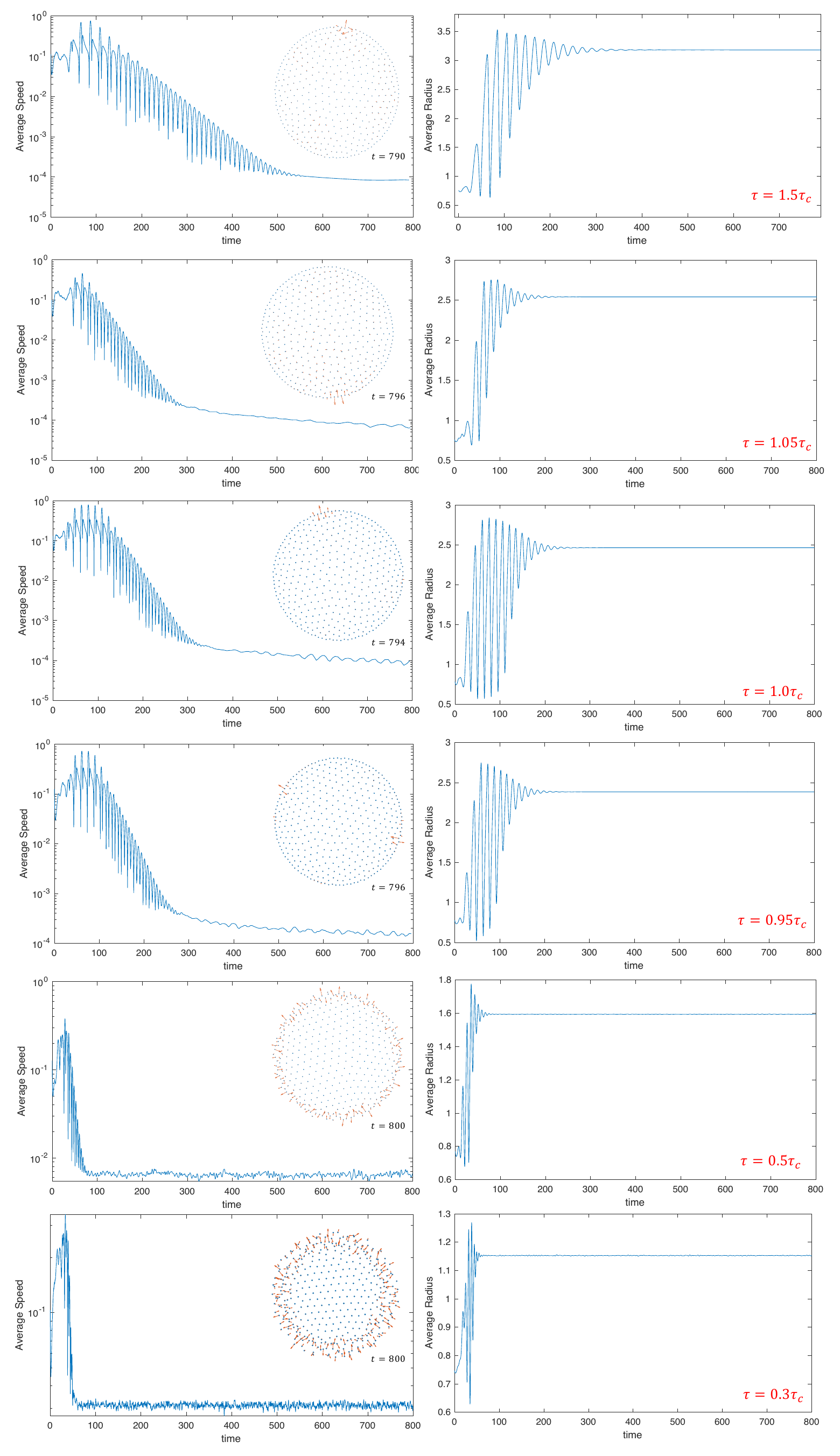}
\caption{Evolution of particle dynamics with decreasing $\tau$.  The boiling surface appears around $\tau_c$, and its width grows as $\tau$ is progressively decreased.  Clusters for $\tau$ below $\tau_l$ are not shown.  Here $N=400$, $J=1$, $K=-0.7$.}
\label{fig:moustail_evolution}
\end{figure}
\end{appendix}

\newpage
\bibliographystyle{apsrev}
\bibliography{ref.bib}    
    
\end{document}